\documentclass[prb,aps,twocolumn]{revtex4}

\usepackage{natbib}
\usepackage{graphicx}
\usepackage{amsmath}
\usepackage{epstopdf}
\usepackage{amsbsy}
\usepackage{color}
\usepackage{dcolumn}
\usepackage{bm}

\newcommand{\ra}{\rangle}
\newcommand{\la}{\langle}

\newcommand{\mb}{\mathbf}

\newcommand{\vk}{{\mathbf{k}}}

\newcommand{\vp}{\mathbf{p}}

\newcommand{\vrr}{\mathbf{r}}
\newcommand{\vR}{\mathbf{R}}

\newcommand{\beq}{\begin{equation}}
\newcommand{\eeq}{\end{equation}}
\newcommand{\bea}{\begin{eqnarray}}
\newcommand{\eea}{\end{eqnarray}}

\begin{document}

\title{Fermionization of Bosons in a Flat Band}

\author{Saurabh Maiti and Tigran Sedrakyan}

\affiliation {Department of Physics, University of Massachusetts,
Amherst, MA 01003}

\date{\today}

\begin{abstract}
Strongly interacting bosons that live in a lattice with degeneracy
in its lowest energy band experience frustration that can prevent
the formation of a Bose-Einstein condensate. Such systems form an
ideal playground to investigate spin-liquid behavior. We use the
variational principle and the Chern-Simons technique of
fermionization of hard-core bosons on Kagome
lattice to find that below lattice filling fraction $\nu=1/3$ the system favors a topologically ordered chiral spin-liquid state
that is gapped in bulk, spontaneously breaks Time-Reversal Symmetry, and supports massless chiral bosonic edge mode. 
We construct the many-body variational wave function of the state and show that the corresponding energy coincides with the energy of the flat band. This result proves that the ground state of the system cannot stabilize a Bose condensate below 
$\nu=1/3$. 
The fermionization and variational scheme we outline apply to any
non-Bravais lattice.  We distinguish between the roles played by the
Chern-Simons gauge field in lattices with a flat band and those
exhibiting a moat-like dispersion (which is degenerate along a closed
contour in the reciprocal space). We also suggest experimental
probes to differentiate the proposed ground state from a condensate.
\end{abstract}


\maketitle

\section{Introduction}
A quantum spin-liquid is one of the sought after states in a
strongly interacting spin system. A recent neutron scattering
experiment on herbertsmithite\cite{Han2012} reported the first
detection of a spin-liquid phase. While the characterization of
this state has been a matter of debate\cite{Punk2014}, this
experiment indicates that the detection of the various spin-liquid
states is not far away. Some manifestations of this state include
a gapless Dirac (4-spinor) spin liquid state coupled to a U(1)
gauge field\cite{Ran2007,Iqbal2011}, a gapped $\mathrm{Z}_2$ spin
liquid state\cite{Yan2011,Punk2014}, a chiral spin liquid (CSL)
\cite{Yang1993,Messio2012,He2014,Gong2014,Bauer2014,Kumar2014,Kumar2015,Zalatel2016}
some of which can also be gapless\cite{Pereira2018}. Such states
exhibit absence of rotational symmetry breaking and, as such, do
not stabilize any long-range magnetic order. Their collective
low-energy excitations support fractionalized statistics, which
can be classified using topological quantum field theory with
various symmetry properties. Variety of techniques have been used
in the literature to identify and study the properties of such
states\cite{Kalmeyer1987,Wen1989,KHVESHCHENKO1989,KHVESHCHENKO1990}
with many of the early and current attempts focusing on 2D
triangular and honeycomb
lattices\cite{KITAEV2006,Yao2007,Xu2009,Wietek2017,Werman2018}.

Amongst the numerous quantum spin-liquid
candidates\cite{Balents2010,Savary2017,Zhou2017,Benton2018}, the
spin-$1/2$ Heisenberg magnet on a Kagome lattice stands out as a
fascinating system that is believed to give rise to a variety of
spin-liquid phases\cite{Norman2016,Pohle2017}. The Kagome lattice is known to posses a flat band (quenched dispersion). If the lattice is
sparsely populated by strongly interacting bosons (also referred
to as hard-core bosons which avoid multiple occupancy of a single site), the state of the system is determined entirely from minimization of the interactions, since the kinetic energy of the system is fully quenched. Such a system is equivalent to an XY model with the
$z$-directional magnetic field term, $H_{mag}=\sum_{\bf r}\mu
S_{\bf r}^z$. The field strength $\mu$ maps on to the chemical
potential of hard-core bosons. These bosons at low densities do not condense
because of the degeneracy of the condensate wave functions which
arises from the flat band. In the XY model,
the absence of condensation translates to the absence of magnetic order. One is thus
interested in learning about phases that can be stabilized in such
a system. 

{ It is instructive to note that if one replaces hard-core bosons by spinless fermions, the flat band would be capable accommodating fermionic states up to $\nu=1/3$, such that
fermions avoid each other and have exactly zero energy (measured relative to the flat band) just by filling the flat band.
This observation suggests that if there was a way for a system of hard-core bosons to stabilize low-energy excitations with fermionic statistics, such a state could be energetically favorable.}
In this article, we demonstrate the use of a
technique that fermionizes hard-core bosons to find a chiral
spin-liquid state as the energetically favorable candidate for the
ground state of interacting spins on a Kagome lattice, which
spontaneously breaks time-reversal symmetry (TRS) and represents an example where topological ordering is realized with interacting bosons.

\begin{figure}[htp]
\includegraphics[width=0.95\columnwidth]{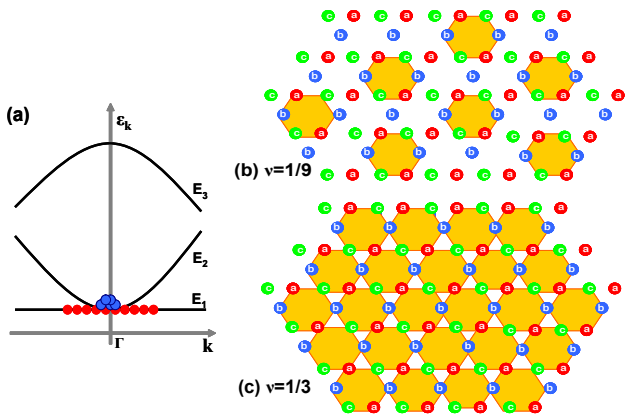}
\caption{\label{fig:BS} (a) The electronic structure of
Kagome-lattice has three bands, with the lowest band ($E_1$) being
flat. The red and blue dots denote the population of $E_1$ and
$E_2$ bands. (b) The real space depiction of the $\nu=1/9$ state.
The shaded region denotes occupancy by a boson. The wavefunctions
do not overlap. (c) The real space depiction of the $\nu=1/3$
chiral spin-liquid state obtained from fermionization.}
\end{figure}

Current understanding of the system under consideration is that the hard-core
bosons can avoid paying any cost of interaction energy by forming
spatially separated localized states\cite{Bergman2008,Huber2010}
which is possible due to the presence of a flat band in the Kagome
lattice. Such a state can persist up to lattice filling of
$\nu=1/9$, beyond which the system is faced with a choice between
(a) populating higher energy bands (see Fig. \ref{fig:BS}); and
(b) letting the bosons still reside in the flat band and paying
the interaction cost due to overlap. The choice (a) would result
in condensation of bosons (represented by (blue) dots on $E_2$ in
Fig. \ref{fig:BS}) to the $\Gamma$ point of the Brillouin zone,
leading to the supersolid state whose chemical potential grows as
$\mu\sim (\nu-1/9)$, up to logarithmic prefactors. Such
a supersolid state has been predicted as a mean-field theory for
weakly interacting bosons at lattice fillings above $1/9$ in
Ref.~\onlinecite{Huber2010}. The corresponding ground state energy $E_{GS}$ scales as $E_{GS}\sim (\nu-1/9)^2$.

The choice (b) essentially remains unexplored. We 
find within our approach that for strongly interacting bosons in a flat band Kagome lattice
the correlations lead to effective fermionization of the bosons. To this end, we show that the system can still save energy
(retaining $E_{\rm GS}\sim0$) by continuing to populate the flat band up to $\nu=1/3$. Interestingly enough, the scaling of the chemical potential of the fermionized system with
particle density is insensitive to the filling fraction around $1/9$, which is a critical value for condensed bosons. This change in the scaling of $\mu$ with density will result in different velocity distribution curves extracted from time-of-flight experiments on trapped atoms. This suggests that such an experiment can be used as a tool to distinguish between the two possible states, (a) and (b), under discussion, above the $1/9$ filling.

The fermionization of hard-core bosons relies on attaching a
Chern-Simons(CS) phase, $\Lambda_{\{\vrr\}}$, to a fermionic many-body
wavefunction: $|\Phi_B\ra=e^{i\Lambda_{\{\vrr\}}}|\Psi_F\ra$
(where $\{\vrr\}$ denotes the set of coordinates of the particles,
$\Psi_F$ denotes the fermionic wavefunction, and $\Phi_B$ denotes
the bosonic wavefunction). While this technique was used
extensively to describe fractional quantum Hall(fQH)
states\cite{Jain1989,Lopez1991,Halperin1993,PhysRevLett.62.82}, it has also been
applied to spin-orbit coupled bosons\cite{Sedrakyan2012}, bosons
in honeycomb lattice\cite{Sedrakyan2015_2}, and bosons living on a
moat\cite{Sedrakyan2014}. It has been shown that fermionization
can stabilize topological spin
ordering\cite{Sedrakyan2017,Rui2018}, high-temperature
superconductivity\cite{Sedrakyan2017_12}, and even a chiral
spin-liquid in a moat band\cite{Sedrakyan2015}. In the latter
case, the role of magnetic frustration is mapped to the degeneracy
of the lowest single-particle states of the fermions (the
moat-band).
In this article, we device a scheme to fermionize bosons in a
non-Bravais lattice in general which allows us to apply the
fermionization technique to arbitrary lattices. For demonstration
purposes, we apply our formulation to the case of Kagome lattice.
{ We find that the trial wavefunction we propose, which by
construction describes a CSL, is an
eigenstate of the XY-model within a flux-smearing mean field
approximation (MFA), and has the lowest many-body ground state
energy. Our MFA breaks TRS and hence has non-zero flux per unit
cell. The resulting spectrum has fermionic excitations and flux-flip excitations that can be understood as fractionally charged vortices with fractional self-statistics with angle $\theta =\frac{\pi}{2}$. These are semion excitations (which are examples of abelian anyons) of 
Kalmeyer-Laughlin CSL\cite{Kalmeyer1987}.
Moreover, the flat band prior to the MFA remains flat, but is now also gapped from the rest of the excited states. This provides a posteriori justification for the stability of our MFA. Indeed, using our trail wavefunction  we explicitly demonstrate that the} flux
distribution within the unit cell is such that the flat band of the
Kagome lattice is preserved upon the TRS breaking. This feature is
unique to our prescription and unlike other attempts in literature
to tackle a similar problem\cite{Yang1993,Kumar2014,Kumar2015}. More explicitly, we set-up a trial wavefunction in the continuum
limit (although this limit is not necessary) and
demonstrate that our state has the lowest energy, potentially up
to $\nu=1/3$. 

{It must be emphasized that the approach discussed in this article considers the effect of the CS field after the flux smearing MFA (which is consistent with the Gauss law constraint imposed by the CS terms in the original action). If one starts from the lattice gauge theory of CS field prior to the MFA\cite{Lopez1991,Kumar2014,Kumar2015,Sohal2018}, it is not straightforward to converge to particular flux distribution, and the question of the TRS breaking remains an interesting open question. For this work, we wish to emphasize the usefulness of our variational MFA, in light of the previous results obtained in Refs. \onlinecite{Sedrakyan2012,Sedrakyan2014,Sedrakyan2015,Sedrakyan2017,Sedrakyan2017_12}, which shows that the ground state can have the lowest possible energy from small occupation numbers all the way to the one-third filling of the lattice. Importantly, our result rules out the possibility of condensation of interacting bosons [choice (a) discussed above] in the vicinity of the $\nu=1/9$ filling fraction of the lattice. One can classify the possible CSLs based on symmetries, as done in some recent studies\cite{Bieri2015,Bieri2016}. The comprehensive classification of possible spin-liquid states based on the flux attachment procedure used in the present work is also possible, and is an open important problem that needs to be investigated in the future.

As indicated above, using fermionization technique, the CSL behavior has been suggested in systems with moats: which has a degenerate minima in the single particle spectrum. In this article we consider Kagome lattice which possess a flat band. We will discuss the differences arising in the calculation of many-body
ground state energy in systems exhibiting moats and flat-bands. Finally we shall present the interesting avenues of research this approach motivates for future work.}

The rest of the article is formatted as follows. Section
\ref{sec:fermionization} reviews the fermionization technique in
general and presents a scheme to carry it out in a non-Bravais
lattice (with multiple atoms per unit cell). In Sec.
\ref{sec:KagomeTrial} we discuss the application of the scheme to
Kagome lattice and show that the CSL trial wavefunction has the
lowest ground state energy. Finally in Sec. \ref{sec:conclusion} we summarize
the work and present an outlook for the fermionization technique.
An appendix is included that presents technical details of some
calculations which would have cluttered the presentation in the
main text.

\section{Fermionization in a non-Bravais lattice}\label{sec:fermionization}
{Let us start by summarizing the concept of fermionization.
Consider a generic N-body Hamiltonian (on a Bravais lattice) $\hat
H(\vrr_1,...,\vrr_N)\equiv \hat H(\{\vrr\})$. Let $|\Phi_B\ra$ be
the undetermined many-body bosonic wavefunction (subject to
hard-shell constraint for the particles).} The many-body ground
state energy, $E_{\rm GS}$, for the hard-core bosons can be
written as $E_{\rm GS}=\la\Phi_B|\hat H(\{\vrr\})|\Phi_B\ra$.
{The wavefunction $|\Phi_B\ra$, describing hard core bosons
can be expressed in a fermionic basis as
$|\Phi_B\ra=e^{i\Lambda_{\{\vrr\}}}|\Psi_F\ra$}, where
$\Lambda_{\{\vrr\}}=\kappa\sum_{r'<r}\theta_{\vrr\vrr'}$ {
[$\theta_{\vrr\vrr'}={\rm arg}(\vrr-\vrr')$]}, $\kappa$ is an odd
integer. Thus, \bea
E_{GS}&=&\la\Psi_F|e^{-i\Lambda_{\{\vrr\}}}\hat
H(\{\vrr\})e^{i\Lambda_{\{\vrr\}}}|\Psi_F\ra\nonumber\\
&=&\la\Psi_F|\hat H(\{\vrr\},\mb
A_{\{\vrr\}})|\Psi_F\ra,\label{eq:EGS1}\\
&\approx& \la\Psi_c|\sum_{\{\vrr\}}\hat H^{MFA}(\vrr,\mb
A_{\vrr})|\Psi_c\ra.\label{eq:EGS2}\eea Here the non local vector
potential $\mb
A_{\{\vrr\}}\equiv\partial_{\vrr}\Lambda_{\{\vrr\}}$ enters into the Hamiltonian via covariant derivative
$-i\partial_{\vrr}\rightarrow-i\partial_{\vrr}+\mb A_{\{\vrr\}}$.
This amounts to a magnetic field of $\mb
B(\vrr)\equiv\mb\nabla\times\mb
A_{\{\vrr\}}=\sum_i2\pi\kappa\delta(\vrr-\vrr_i)$. { While,
the wavefunction $|\Psi_F\ra$ in Eq. (\ref{eq:EGS1}) is still
undetermined, Eq (\ref{eq:EGS2}) presents a way to estimate
$E_{GS}$ and needs further explanation. To go from $\hat
H(\{\vrr\},\mb A_{\{\vrr\}})$ to $\hat H^{MFA}(\vrr,\mb
A_{\vrr})$, we perform a flux-smearing MFA. The essence of this
approximation is that  $\mb
B(\vrr)=\sum_i2\pi\kappa\delta(\vrr-\vrr_i)$
$\rightarrow2\pi\kappa\la\hat n(\vrr)\ra=2\pi\kappa\nu$. That is,
the field that was pinned to every particle, is smeared and
treated as uniform. This MFA is consistent with the Gauss law constraint ($B=2\pi\kappa\nu$) imposed by the CS field prior to performing the MFA. 

An important distinction is to be made here. In the continuum limit, at large enough length scales such that $|\vrr|\sim l\gg 1/\sqrt{\rm density}$, the Gauss law constraint can be accounted for by introducing a local $\mb A_{\vrr}$ such that $\mb{\nabla}\times\mb
A_{\vrr}=\mb B$. For the lattice version of this approximation, we require the constraint to be implemented at the level of `flux per unit cell' (hence $B$ is related directly to $\nu$ in the above formulas and not the density). For the case of 1 atom per unit cell with nearest neighbor hoppings, the flux configuration of the CS field is no different from the usual Maxwell field. The wavefunctions $|\Psi_c\ra$ in Eq. (\ref{eq:EGS2}) is then formed from the slater determinant of the single particle states of $\hat
H(\vrr,\mb A_{\vrr})$.}

\begin{figure}[htp]
\includegraphics[width=0.9\columnwidth]{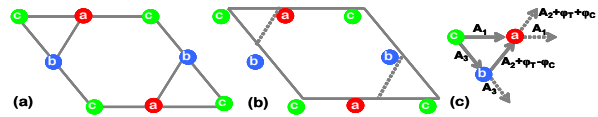}
\caption{\label{fig:2} (a) A unit cell of the Kagome lattice. (b)
The unit cell redrawn with a shift. This explicitly shows that the
cell area includes 3 atoms. The dashed lines are the internal
bonds chosen to not include any atoms in the triangular loop. All
the particles, and hence the flux, is contained within the
hexagon. (c) Flux attachment within the unit cell of the Kagome
lattice. $A_i$ denotes the phase accumulated by $\mb A(\vrr)$
while traversing the direction $\mb a_i$ such that $A_1=A_2+A_3$.}
\end{figure}

\begin{figure}[htp]
\includegraphics[width=0.9\columnwidth]{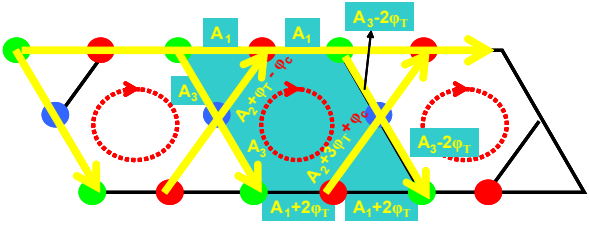}
\caption{\label{fig:3} Extended scheme for the flux attachment.
Note that $\phi_T$ grows to account for the area law, where as
$\phi_C$ is the same in every unit cell. The flux through the
hexagon and the unit cell is $8\phi_T$ and through any triangle is
$\phi_T-\phi_C$. In our MFA $\phi_T=\phi_C$.}
\end{figure}

For a non-Bravais lattice, {neither the nature of flux
smearing} MFA, nor the construction of a variational many-body
state is straightforward. {When there are two (or more) atoms
per unit cell, had we carried out the flux smearing approximation
as for a Bravias lattice, there would be no distinction between
the Maxwell-type field and CS field. We realize that small enough length scales $|\vrr|\sim$lattice constant, the flux distribution within a unit cell must depend on the relative locations of atoms within the unit cell. Thus it is possible to implement the Gauss-law constraint at the level of unit cell and still be left with a degree of freedom in distributing the flux internally in the unit cell. To implement this feature, we propose that this can
be implemented by requiring that all the hops that form loops
internal to the unit cell (not more that one sharing edge), must
not enclose any flux. This is based on the observation that in the absence of CDW ordering, the homogeneous 
CS flux attachment preserves the fact that if there is a triangle composed of two nearest-neighboring and one next-nearest-neighboring sites, then there must be such triangle within the unit cell where two subsequent hops along the links of nearest-neighboring sites are equivalent to a single hop along the next-nearest-neighbor link, see approaches in Refs. \onlinecite{Sedrakyan2012,Sedrakyan2014,Sedrakyan2015,Sedrakyan2017,Maiti2019}.

Note that if we have only
one atom per unit cell (Bravais lattice), then there are no internal loops possible and thus the CS flux is the same as Maxwell flux. However, when we consider a non-Bravais lattice which contains loops of hops
internal to the unit cell, there can be parts of the unit cell with
zero net flux\cite{Sedrakyan2014, Sedrakyan2015_2}.} This is seen by redrawing
the unit cell as in Fig. \ref{fig:2}b. { The flux smearing field that confirms with the Gauss law can be accounted for by $\mb A_{\vrr}$ which introduces a
flux $\phi_T$ per triangular region of the Kagome lattice.} To
account for the zero flux regions, one must introduce an
intra-unit cell flux $\phi_C=\phi_T$ as in Figs.
\ref{fig:2}c,\ref{fig:3}. This introduction is lattice dependent and will be demonstrated later for the Kagome lattice.

{Having demonstrated the non-triviality in the flux smearing
MFA, we now address the subtlety in constructing the many-body
wavefunction}. The Hamiltonian matrix in a non-Bravais lattice has
a rank, $n$, that is the number of atoms in the basis of the
lattice. An $N$-body wavefunction can be denoted as
$\Psi^{\{a\}}_{\{n\}}(\{\vrr\})\equiv\Psi_{n_1,n_2,...,n_N}^{a_1,a_2,...,a_N}(\vrr_1,\vrr_2,...,\vrr_N)$,
where a given coordinate $\vrr_i$ can describe the wavefunction
component $a_i\in\{1,2,...,n\}$ in quantum state $n_i$.
Correspondingly, the N-body Hamiltonian acquires the form $\hat
H_{\{ab\}}(\{\vrr\})\equiv$\beq\label{eq:NBH} \hat
H_{a_1b_1,...,a_Nb_N}(\vrr_1,...,\vrr_N)=\sum_{i=1}^N \hat
H_{a_ib_i}(\vrr_i)\prod_{j\neq i}^N\delta_{a_jb_j}. \eeq The
presence of a basis in a non-Bravais lattice introduces a degree
of freedom in implementing the antisymmetrization: { if the
single particle states are given by $\psi^a_{n}(\vrr)$, where $n$
denotes the band and $a$ denotes the component of the
wavefunction, the most general construction of a fermionic N-body
state is \bea\label{eq:MBWF}
\Psi_{\{s\}}(\{\vrr\})&=&M^{s_1s_2...s_N}_{\{s\}}\times\nonumber\\
&&\frac{1}{\sqrt{N!}}\rm{Det}\left[\psi^{s_1,s_2,...,s_n}_{n_1,n_2,...,n_N}(\vrr_1,\vrr_2,...,\vrr_N)\right],\nonumber\\
\eea where $s,s_i\in\{1,2,...,N\}$, the repeated indices are
summed over, and the Slater determinant is formed out of the
indices $n_i$ and $\vrr_i$. The tensor $M$ superposes the
anti-symmetrized Slater determinants for various combinations of
$\{s_i\}$. The constraints on $M$ are enforced by requiring the
many-body wavefunction be normalized to unity, and that the
permutation properties of $M$ must respect antisymmetrization.
This formally sets up a variational problem where the choice of
the $M$-tensor is variational. The problem of such a huge
variational space can be overcome by resorting to another
technique to guess a trial wave-function, which we discuss next.}

\subsection{Trial wavefunction by projection to a band}
We note that a single particle state is indexed as
$\psi^{a}_{n,\vk}(\vrr)$. Indices $a$ and $n$ are necessary to
account for the non-Bravais nature of the lattice. Index $\vk$
reflects the crystal translational symmetry which is independent
of the non-Bravais nature and allows us to write (see
Appendix C)
\beq\label{eq:psi}\psi^a_{n,\vk}(\vrr)=\int_{\vrr'}R^a_n(\vrr-\vrr')\phi_{\vk}(\vrr'),\eeq
where $\phi_{\vk}(\vrr)$ is a solution to the characteristic
equation of $\hat H_{ab}(\vrr)$, and $R_n^a(\vrr)$ is the Fourier
transform of the eigenvectors of $H_{ab}(\vk)$. The normalization
condition is enforced by requiring
$\int_{\vrr_1,\vrr_2,\vrr}\sum_a
\phi^*_{\vk}(\vrr_2)R^{a*}_n(\vrr-\vrr_1)R^{a}_n(\vrr-\vrr_2)\phi_{\vk}(\vrr_2)=1$.
The many-body wavefunction can then be constructed
as\beq\label{eq:mb}
\Psi^{\{a\}}_{\{n,\vk\}}(\{\vrr\})=\int_{\{\vrr'\}}\!\!\!R^{a_1}_{n_1}(\vrr_1-\vrr'_1)\ldots R^{a_N}_{n_N}(\vrr_N-\vrr'_N)\phi_{\{\vk\}}(\{\vrr'\}).
\eeq Here $\phi_{\{\vk\}}(\{\vrr\})$ denotes the N-body
wavefunction formed out of the quantum states $\{\vk\}$ and
coordinates $\{\vrr\}$. The energy expectation value of band $n$
and quantum states $\{\vk\}$ is given by\beq\label{eq:energy}
E_n(\{\vk\})=\sum_{\{a\},\{b\}}\int_{\{\vrr\}}
\Psi^{\{a\}*}_{\{n,\vk\}}(\{\vrr\})\hat
H_{\{ab\}}(\{\vrr\})\Psi^{\{b\}}_{\{n,\vk\}}(\{\vrr\}). \eeq Using
Eq. (\ref{eq:NBH}), together with the normalization condition, we
can show that \beq\label{eq:ham0}
E_{n\{\vk\}}=\int_{\{\vrr'\}\{\vrr''\}}\phi^{*}_{\{\vk\}}(\{\vrr'\})E_n(\{\vrr'\},\{\vrr''\})\phi_{\{\vk\}}(\{\vrr''\}),
\eeq 
where we introduced a notation
\bea\label{eq:ham1}
E_n(\{\vrr'\},\{\vrr''\})&=&\sum_{\{a\}\{b\}}\int_{\{\vrr\}}\prod_i R^{*a_i}_n(\vrr'_i-\vrr_i)\nonumber\\
&&~\times\hat
H_{\{ab\}}(\{\vrr\})\prod_jR^{b_j}_n(\vrr_j-\vrr''_j). \eea We
have thus devised a way to remove the index dependence of $\hat
H_{ab}(\vrr)$ and map it to a single component energy function
$E_n(\vrr',\vrr'')$ with single component wavefunction
$\phi_{\vk}(\vrr)$. { The advantage of doing this is that we
can readily use Eqs (\ref{eq:EGS1}) and (\ref{eq:EGS2}) without
resorting to multicomponent nature of $H$, which may lead to
introduction of a non-Abelian CS field.} The many body state
$\phi_{\{\vk\}}(\{\vrr\})$ has to be bosonic. But it can be
fermionized as: \bea\label{eq:ham2}
\phi_{\{\vk\}}(\{\vrr\})&=&e^{i\Lambda_{\{\vrr\}}}\psi_{\{\vk\}}(\{\vrr\}),
\eea where $\psi_{\{\vk\}}(\{\vrr\})$ is a Slater determinant.
Thus the fermionized version of Eq. (\ref{eq:ham0}) can be
achieved by promoting\bea\label{eq:tr}
\phi_{\{\vk\}}(\{\vrr\})&\rightarrow&\psi_{\{\vk\}}(\{\vrr\}),~\text{and}\nonumber\\
E_n(\{\vrr'\},\{\vrr''\})&\rightarrow&
e^{-i\Lambda_{\{\vrr'\}}}E_n(\{\vrr'\},\{\vrr''\})e^{i\Lambda_{\{\vrr''\}}}.
\eea We note that while $\hat H(\{\vrr\})$ is entirely the
property of the underlying lattice, the construction of $E_n$ and
the choice of $\psi$ is a variational knob available to us. {
Thus we have traded the M-tensor based variational space with the
choice of $E_n(\vrr,\vrr')$}. In what follows, we demonstrate that
a trial wavefunction for hard-core bosons on Kagome lattice,
derived from the above scheme, describes a CSL with spontaneously
broken TRS and has the lowest possible $E_{GS}$.

\section{Kagome Lattice and the many-body trial wavefunction}\label{sec:KagomeTrial}
{In general, the problem of hard core bosons on a lattice can
be studied by looking at the spin-1/2 XY model with
Hamiltonian\beq\label{eq:1}
H=\sum_mJ_m\sum_{\vrr,n}S^+_{\vrr}S^-_{\vrr+\vrr_{mn}}~+~\rm{h.c.}
\eeq Here $S^{\pm}=S_x\pm iS_y$ are the spin-1/2 operators; the
index $n$ scans all the neighbors at distance $m$; $\vrr_{m,n}$ is
the vector to the $(m,n)^{\rm th}$ nearest neighbor. The choice of
lattice is reflected in the choice various $\vrr_{mn}$. One choice
of the phase attachment(in second-quantized notation) that
accomplishes the CS transformation
is\cite{Jain1989,Lopez1991,Halperin1993} \beq\label{eq:2}
S^{+}_{\vrr}=c^{\dag}_{\vrr}e^{i
\Lambda_{\vrr}},~~S^{-}_{\vrr}=e^{-i\Lambda_{\vrr}}c_{\vrr},\eeq
where $c^{\dag}_{\vrr}$ is a fermionic creation operator, and
\beq\label{eq:3}\Lambda_{\vrr}\equiv
\kappa\sum_{\vrr\neq\vrr'}\theta_{rr'}c^{\dag}_{\vrr'}
c_{\vrr'}.\eeq  This transforms the Hamiltonian to a fermionic
basis: \beq\label{eq:4}
H=\sum_mJ_m\sum_{\vrr,n}c^{\dag}_{\vrr}e^{i\Lambda_{\vrr,\vrr+\vrr_{mn}}}
c_{\vrr+\vrr_{mn}} ~+~\rm{h.c.},\eeq where
$\Lambda_{\vrr_1,\vrr_2}\equiv\Lambda_{\vrr_1}-\Lambda_{\vrr_2}$
evaluated along the line joining $\vrr_1$ and $\vrr_2$. It is the
analog of the accumulated phase $\int^{\vrr_2}_{\vrr_1}\mb A\cdot
d\mb l$ in a lattice. Geometrically, $\Lambda_{\vrr_1,\vrr_2}$ is
the sum of the angles subtended by the vector $\vrr_1-\vrr_2$ at
every other site (located at $\vrr'$), weighted by the occupation
probability of all sites along the path
$\vrr_1\rightarrow\vrr_2$.}

From here on we specialize to the Kagome lattice with the first
neighbor hoppings. This is achieved by setting $m=1$, and letting
$n$ scan from $1$ through $4$ (4 nearest neighbors) for each of
the three atoms in the unit cell. The MFA leads to
$c^{\dag}_{\vrr}c_{\vrr}\rightarrow\la
c^{\dag}_{\vrr}c_{\vrr}\ra=n_{\vrr}=\nu$ (the filling fraction in
the lattice) in the definition of $\Lambda_{\vrr}$. Let the
lattice now be populated by hard-core bosons at every site with
filling fraction $\nu$. Within our MFA, this provides non-zero
flux at each site, spontaneously breaking TRS. The flux
distribution is such that all the flux ($3\times2\pi\nu\kappa$) is
concentrated through the hexagon (Fig \ref{fig:3}). To achieve
this, one has to introduce two fluxes $\phi_T$ [to account for the
Maxwellian field $\mb A_{\vrr}$], and $\phi_C$ (to account for the
the intra unit cell flux modulation). In the absence of any
external field, the CS field requires
$\phi_C=\phi_T=3\pi\nu\kappa/4$.

\begin{figure}[htp]
\includegraphics[width=.99\columnwidth]{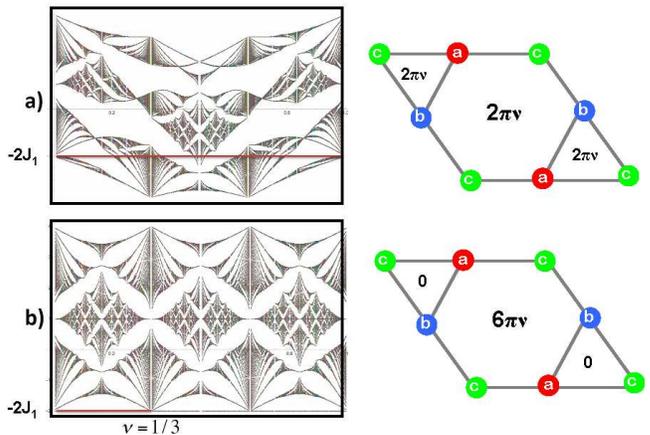}
\caption{\label{fig:4} Comparison of the Kagome energy spectrum in
a CS field for two different flux distributions (a) and (b). The
flat band is preserved in (b) where the flux through the triangle
is zero. In this case, the spectrum is unique only up to
$\nu=1/3$.}
\end{figure}

We take note of the fact that the single particle
`Hofstadter' spectrum of this system is sensitive to the flux
distribution within the unit cell. Figure \ref{fig:4}a,b
demonstrates that different flux distributions lead to different
spectra. The property of the CS flux distribution seems to be that
(i) the spectrum is unique up to $\nu=1/3$ at which point the flux
through the unit cell is $2\pi$; (ii) the lowest energy band is
still flat! We thus observe that a CS-flux distribution in the
Kagome unit cell leads to an isolated flat band.

\begin{figure}[htp]
\includegraphics[width=.99\columnwidth]{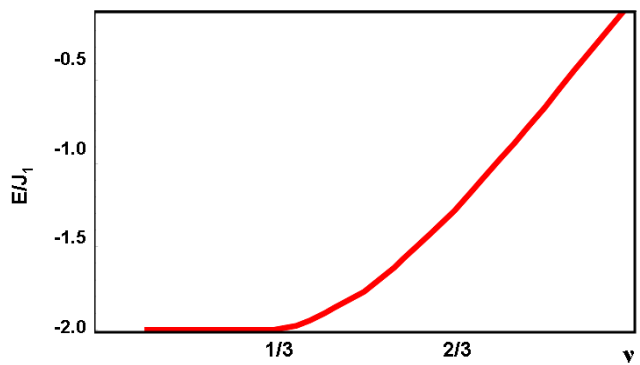}
\caption{\label{fig:5} Ground state energy of non-interacting
fermions in Kagome lattice subject to the CS flux at mean-field
level. Energy begins to rise after $\nu=1/3$ filling.}
\end{figure}

{We now resort to calculating $E_{GS}$ where we will need the
many-body trial wavefunction. To be explicit, we shall demonstrate
this in the continuum limit. We iterate that to calculate $E_{GS}$
for the original Hamiltonian, we make an estimate using a trial
wavefunction which is bosonic, but constructed out of a
(fermionic) slater determinant by attaching a CS phase. The single
particle states needed to construct this slater determinant are
the eigenstates of the Hamiltonian obtained after the
flux-smearing approximation. Further, since the original
Hamiltonian has three bands, we focus of forming the slater
determinant from the single particle states of the band in which
the chemical potential is expected to lie. Since we will be
addressing the low-density regime, we expect the chemical
potential to lie in the lowest band. Using the projection technique
introduced above, we can express $E_{GS}$ in terms of
$E_n(\vrr,\vrr')$, where $n$ corresponds to the lowest flat band [see
Eqs. (\ref{eq:ham0})-(\ref{eq:tr})].}

We choose $R^a_n$ and $\psi$ to be the eigenstates of the $\hat
H^{\rm MFA}(\vrr,\mb A_{\vrr})$. This estimate for $E_{GS}$ is
expected to account for the non-local nature of $\mb A(\{\vrr\})$
and provide corrections to the many-body energy computed within
MFA (Fig. \ref{fig:5}). We shall explicitly demonstrate this in
the continuum limit where the Hamiltonian in Eq. (\ref{eq:1})
after flux smearing can be written as:
\beq\label{eq:H}\hat{H}^{\rm MFA}(\vrr)=J_1\left(
\begin{array}{ccc}
0&H_2&H_1\\
H^*_2&0&H_3\\
H^*_1&H^*_3&0
\end{array}\right),\eeq where $H_i\equiv \left(e^{i\vp\cdot\mb a_i/2}+e^{-i\vp\cdot\mb a_i/2}\right),~i\in\{1,3\}$; $H_2\equiv \left(e^{i\vp\cdot\mb a_2/2}+e^{-i\vp\cdot\mb a_2/2}\right)e^{i\phi_C}$; $\vp\equiv-i\partial_{\vrr}+\mb A_{\vrr}$, $p_i\equiv\mb a_i\cdot\vp$; and $\mb a_1 =
a(1,0);~\mb a_2 = a(1,\sqrt{3})/2;~\mb a_3 =\mb a_1-\mb a_2$ are
the translation vectors. It is understood that the Hamiltonian is
to be expanded to $\mathcal{O}(p^2)$. The resulting characteristic
equation is \beq\label{eq:char2}
\frac{E}{J_1}\left(\frac{E}{J_1}+2\right)\left[\left(\frac{E}{J_1}+2\right)\left(\frac{E}{J_1}-4\right)+\frac32\vp^2\right]\psi(\vrr)=0.
\eeq It can be shown that $E=0$ is a trivial solution. The
eigenvalues are thus $E_1=-2J_1$,
$E_{2/3}=J_1\left(1\mp\sqrt{9-3Ba^2(m+1/2)}\right)$,
$m\in\{0,1,2,...\}$, and $\mb B=\mb\nabla\times\mb A_{\vrr}$. Note
that the flat band $E_1$ is gapped from the rest of the dispersing
bands($E_{2,3}$) by $3J_1(1-\sqrt{1-Ba^2/6})$. From Eq.
(\ref{eq:H}), we note that the wavefunction components
$\psi^i\equiv\int R^i\psi$ ($i\in{a,b,c}$) for $E_1$ satisfy \bea\label{eq:CSflat}
[-p_2^2]\psi^a+[p_2p_3-i(\phi_T-\phi_C)]\psi^c&=&0,\nonumber\\
\left[p_3
p_2+i(\phi_T-\phi_C)\right]\psi^a+[-p_3^2]\psi^c&=&0,\eea where
$\phi_T\equiv\frac{\sqrt{3}}{16}Ba^2$ (measured counterclockwise).
In the absence of any external magnetic field, the CS flux
constraint requires $\phi_T=\phi_C$. Up to $\mathcal{O}(p^2)$, we
can also show that the wavefunction corresponding to $E_1$ and in
quantum state $i$ is (see Appendix D) for calculation
of $\mathcal{N}$): \beq\label{eq:flat band4}
\psi_i(\vrr)=\frac{1}{\mathcal{N}}\left(\begin{array}{c}
p_3\\
-p_1\\
p_2\end{array}\right)f_i(\vrr), \eeq where $\mathcal{N}$ is a
normalization constant and  $f_i(\vrr)$ can be any function that decays
stronger than $1/r^2$ (for normalizability). Similar to the Landau
problem for free electrons, flat band wavefunctions in the
presence of CS field are also inherently localized. To impose
analyticity of the wavefunction, we postulate
$f_i(\vrr)=e^{-(\vrr-\vR_i)^2/2l^2_{CS}}$, where $\vR_i$ is the
center of the localized state, and $l_{CS}$ is an undermined
localization length scale in the theory.

\subsection{Ground-state energy beyond single particle spectrum}
{The energy profile with respect to the filling fraction at
the single particle level is provided in Fig. \ref{fig:5}.} Having
found the wavefunction of a gapped isolated flat band, we can use
Eqs. (\ref{eq:ham0})-(\ref{eq:tr}) with $n=1$ for low enough
density. We observe that because of the non-dispersing nature of
the isolated flat band, the effect of the non-local $A_{\{\vrr\}}$
drops out and the many-body ground state energy is still the same
$(E_{GS}=-2J_1)$ as {computed within the single particle
picture. In other words, there is no statistical correction to the
$E_{GS}$ estimated from the single particle spectrum. This is,
however, only true up to $\nu=1/3\kappa$. This is special case of
flat bands and is not} true for other cases of
fermionization\cite{Sedrakyan2012}, e.g. when there is a moat
band. {In the case of a moat, the statistical
correction to $E_{GS}$ from the flux attachment procedure leads to a scaling\cite{Sedrakyan2012}
$E_{GS}\sim\mu^2\ln^2\mu$ which is still lower than other
proposals for the corresponding ground state without
Fermionization.}

{The reason for the statistical correction to exist for the
moat and not for the flat band can be attributed to the fact that a moat
is degenerate along a 1-D manifold. This means that any finite
$\mu$ requires populating energy levels away from the moat levels
(which have the minimum energy). In an isolated flat band, this
scenario does not arise. Residual interactions between bosons (or
the corresponding fermions) will lead to energy corrections but
the details depend on modeling the interaction matrix element and
is beyond the scope of this work.}

Returning back to the lattice problem, { we may ask up to
what filling does the statistical correction remain zero? This can
be answered by} noting that attaching a flux of $p/q$ to a unit
cell causes the BZ to fold over $q$-times. We have proven that a
CS flux distribution ($p\neq0$) retains the flat band of the
system at $p=0$. A remarkable consequence of this is that for any
$q$ the degeneracy of the flat band is always the same as the
system with no flux attachment. {This can only be split by
residual interactions}. For a Kagome lattice with $N$ unit cells
($3N$ atoms), $N$ states correspond to the flat band. Even though
the flux attachment changes with $\nu$, we can now conclude that
the flat band can remain occupied up to $\nu=1/3$ (for $\kappa=1$). This state is depicted in Fig .\ref{fig:BS}c. {Thus we can
rigorously state that the statistical correction is absent up to
$\nu=1/3$}.

{In the absence of residual interactions, the contribution to
the many-body ground state only comes from the lowest energy of
the single particle spectrum. Since we have demonstrated that, up
to $\nu=1/3$ (for $\kappa=1$), the statistical correction (which
has to be positive definite) is zero, this has to be the minimum
energy ground state.}

We take note of the fact that the spinless fermionic description,
where no spatial symmetry is broken, naturally implies lack of any
spin-ordering ordering in the language of hard-core bosons. Thus,
fermionization is a natural tool that can be used to describe a
spin-liquid state in strongly interacting bosons. Since our MFA breaks TRS, we expect the spin-liquid state to
be chiral. At this stage, we are able to conclude that strongly
interacting bosons in a Kagome lattice favor a chiral spin-liquid
state that spontaneously breaks TRS.

Further, the flux modulation in Figs.\ref{fig:3} and \ref{fig:4}b
actually corresponds to a Chern insulator with staggered flux
$\phi_{C}=3\pi\kappa/4$ threading corner equilateral triangles of
the unit cell with $-2\phi_{C}$ threading the hexagon, superposed
with a uniform external flux of $8\phi_T=6\pi\nu\kappa$ per unit
cell. Any finite $\phi_{C}$ opens a gap at the band-touching
points and defines Chern numbers for each of the three bands. The
lowest band, in this case, will have a Chern number $C=1$, which
cannot be altered unless the system undergoes a phase transition
with the closing of the gap. Thus the field theory of chiral spin-liquid outlined above can be regarded as a theory topologically nontrivial Chern insulator coupled to the
fluctuating Chern-Simons gauge field. Because of the topological
nature of the Chern insulator, Fermion fluctuations here will give
rise to an additional Chern-Simons term in the low-energy
effective action giving a Chern-Simons theory defined by a ``K
matrix" with $K=2$~~\cite{Lu2012}. This implies that the vortex
excitations in this system have fractional
statistics\cite{Wen1995} with statistical angle $\theta = \pi/2$
corresponding to semions.

\section{Summary and outlook}\label{sec:conclusion}
We prescribed a general scheme to construct the N-body
wavefunction and compute the ground state energy of hard-core
bosons in a non-Bravais lattice using fermionization. Using the
example of the Kagome lattice, we showed that a CS type flux
attachment can retain the massive degeneracy of system's original
electronic structure. {Our trial wavefunction suggests that
the ground state of hard-core bosons on a Kagome lattice is a
spontaneously TRS broken chiral spin-liquid state.} We proved
{that our trial wavefunction has zero statistical correction
to the ground state energy due to a 2-dimensional degeneracy of
the flat band in a Kagome lattice. It is thus a lowest energy state that implies absence of condensation of hard core bosons below third-filling (including the vicinity of the $\nu=1/9$ lattice filling fraction discussed above). While within our analysis, it is not possible to
determine the uniqueness of this spin-liquid ground state, corroboration with
other numerical techniques (e.g. DMRG) can help confirm this
state.}

The lattice itself can be realized using a cold atom set-up as in
Ref. \onlinecite{Jo2012}. In this reference, to ensure that the
flat band is the lowest of the three, one can tune the lattice to
the frustrated hopping regime with the help of artificial gauge
fields attaching $\pi$ phases to all links resulting in the
flipped sign of the matrix element. Other verifiable properties are bosonic edge states
(can be detected using sudden decoupling
technique\cite{Atala2014}); and fractional excitations
(time-of-flight experiments and Bragg
spectroscopy\cite{Stenger1999,Rey2005,Ernst2009}).\\

\textit{Acknowledgements}: We are grateful to A. Kamenev and L. I.
Glazman for useful discussions. T.A.S. thanks the Aspen Center for
Physics (supported by National Science Foundation grant
PHY-1607611) and the Max Planck Institute for the Physics of
Complex Systems for hospitality. The work is supported by startup funds
from UMass Amherst.

\appendix
\section{Momentum space to real space wavefunctions}
Prior to implementing the MFA, we quickly review the Kagome
Hamiltonian at the single particle level and find the
wavefunctions of the flat band in the continuum limit. This will
set us up to tackle the scenario with the CS flux distribution.
The lattice Hamiltonian from Eq. (\ref{eq:4}), without $\Lambda$
can be written in $\vk$-space as \beq\label{eq:Ham}
H=\sum_{\vk}\bar\Psi^{\dag}_{\vk}\bar{\mathcal{H}}_{\vk}\bar\Psi_{\vk},\eeq
where $\bar\Psi^{\dag}_{\vk}=(\bar c^{\dag}_{a,\vk},\bar
c^{\dag}_{b,\vk},\bar c^{\dag}_{c,\vk})$. The annihilation
operators are given by
$$\bar c_{x,\vrr}=\sum_{\vk}\bar c_{x,\vk}{\rm e}^{i\vk\cdot
\vrr},$$ such that $x\in\{a,b,c\}$. The vector $\vrr$ only runs
over lattice translations and not internal bonds. Lastly,
\bea\label{eq:HamK}\bar{\mathcal{H}}_{\vk}&=&J_1\left(
\begin{array}{ccc}
0&(1+{\rm e}^{i\vk\cdot \mb a_2})&(1+{\rm e}^{i\vk\cdot \mb a_1})\\
(1+{\rm e}^{-i\vk\cdot \mb a_2})&0&(1+{\rm e}^{i\vk\cdot \mb a_3})\\
(1+{\rm e}^{-i\vk\cdot \mb a_1})&(1+{\rm e}^{-i\vk\cdot \mb
a_3})&0
\end{array}\right),\nonumber\\\eea
where \beq\label{eq:TransVec} \mb a_1 = a(1,0);~~\mb a_2 =
a\left(\frac12,\frac{\sqrt{3}}{2}\right); ~~\mb a_3 =
a\left(\frac12,-\frac{\sqrt{3}}{2}\right).\nonumber\eeq Note that
in addition to the lattice translation vectors $\mb a_1$ and $\mb
a_2$,  we have introduced $\mb a_3=\mb a_1-\mb a_2$. It will be
useful to perform a gauge transformation:
$\mathcal{H}_{\vk}=M^{\dag}\bar{\mathcal{H}}_{\vk}M$ and
$\Psi_{\vk}=M^{\dag}\bar\Psi_{\vk}$ where
$M^{\dag}=\rm{diag}\left(1,e^{-i\vk\cdot\mb
a_2/2},e^{-i\vk\cdot\mb a_1/2}\right)$ such that
\beq\label{eq:Hs}\mathcal{H}_{\vk}=J_1\left(
\begin{array}{ccc}
0&H_2&H_1\\
H_2&0&H_3\\
H_1&H_3&0
\end{array}\right),\eeq and $H_i\equiv \left(e^{i\vk\cdot\mb a_i/2}+e^{-i\vk\cdot\mb a_i/2}\right)$. The resulting characteristic equation to find the eigenvalues is
\beq\label{eq:char}
\left(\frac{E}{J_1}\right)^3-\frac{E}{J_1}\left(H_1^2+H_1^2+H_3^2\right)-2H_1H_2H_3=0.
\eeq The eigenvalues are $E_1=-2J_1$ and
$E_{2/3}=J_1\left(-1\mp\sqrt{1+H_1H_2H_3}\right)$. Note that $E_1$
is independent of any parameters in the Hamiltonian and hence
dispersionless. The wavefunction corresponding to this flat band
is\beq\label{eq:flat band}
\Psi^1_{\vk}=\frac{1}{N_1}\left(\begin{array}{c}
e^{i(\vk\cdot\mb a_1+\vk\cdot\mb a_2)/2}\sin\frac{\vk\cdot\mb a_3}{2}\\
-e^{i(\vk\cdot\mb a_1-\vk\cdot\mb a_2)/2}\sin\frac{\vk\cdot\mb a_1}{2}\\
e^{-i(\vk\cdot\mb a_1-\vk\cdot\mb a_2)/2}\sin\frac{\vk\cdot\mb
a_2}{2}\end{array}\right), \eeq where
$N^2_1=4\left(\sin^2(\vk\cdot\mb a_1/2)\right)+\sin^2(\vk\cdot\mb
a_2/2)+\sin^2(\vk\cdot\mb a_3/2)$.

\begin{figure}[htp]
$\begin{array}{cc}
\includegraphics[width=0.4\columnwidth]{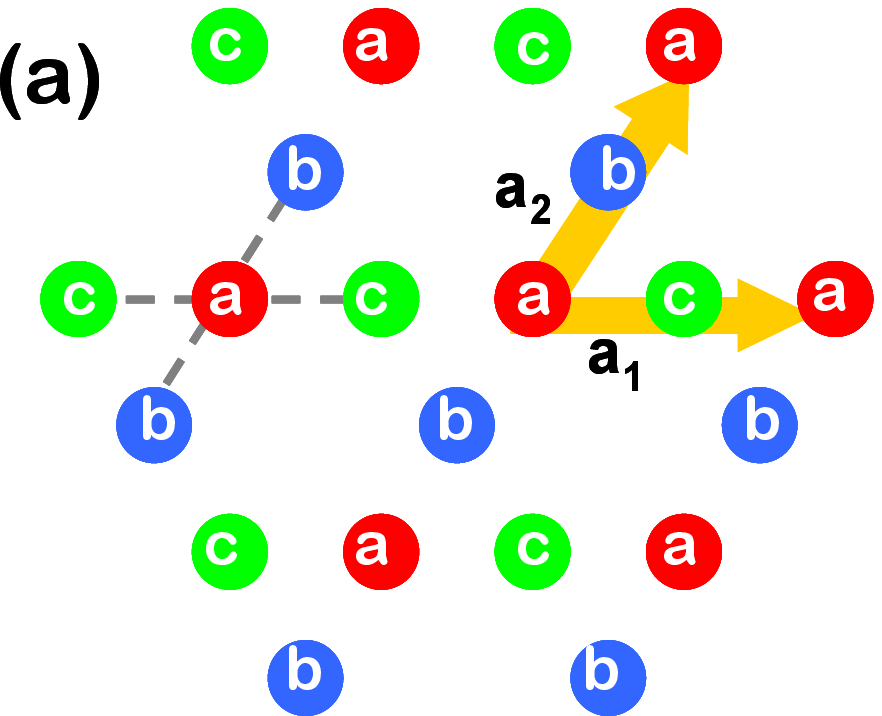}&
\includegraphics[width=0.38\columnwidth]{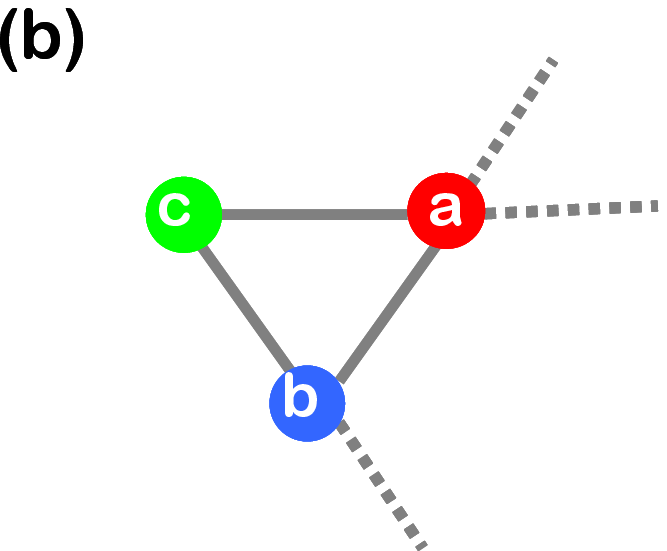}
\end{array}$ \caption{\label{fig:1} (a) The Kagome lattice
has three atoms per unit cell marked $a,b,c$. For each atom, we
consider the first neighbor hoppings ($J_1$) shown by the dashed
lines. $\mb a_{1,2}$ are the lattice translation vectors (b) One
possible choice of the unit cell. All atoms and bonds wholly
belong to the chosen unit cell. The dashed lines are the bonds
that extended to the neighboring unit cells and the solid lines
are the bonds within this unit cell.}
\end{figure}

The continuum limit can be obtained by studying the Hamiltonian
around the $\Gamma$-point. We shall restrict the terms in the
Hamiltonian to $\mathcal{O}(k^2)$. This will result in
$H_i=2-k_i^2$, where $k_i\equiv\vk\cdot\mb a_i$. The resulting
characteristic equation is \beq\label{eq:char2s}
\frac{E}{J_1}\left(\frac{E}{J_1}+2\right)\left[\left(\frac{E}{J_1}+2\right)\left(\frac{E}{J_1}-4\right)+\frac32\vk^2\right]=0.
\eeq It can be shown that $E=0$ is not non-trivial solution. The
eigenvalues are thus $E_1=-2J_1$,
$E_{2/3}=J_1\left(1\mp3\sqrt{1-\vk^2/6}\right)$. And the flat band
wavefunction [up to $\mathcal{O}(k^2)$] is \beq\label{eq:flat
band2} \Psi^1_{\vk}=\sqrt{\frac{2}{3\vk^2}}\left(\begin{array}{c}
k_3\\
-k_1\\
k_2\end{array}\right)=\sqrt{\frac23}\left(\begin{array}{c}
\cos\left(\theta_{\vk}+\frac{\pi}{3}\right)\\
\cos(\theta_{\vk}+\pi)\\
\cos\left(\theta_{\vk}-\frac{\pi}{3}\right)\end{array}\right).
\eeq Note that the flat band wavefunction has the property that a
rotation of $\pi/3$ causes the weights on the sub lattice to
rotate as $a\rightarrow b\rightarrow c\rightarrow a$ and causes
the wavefunction to acquire a phase of $\pi$.Tthe Kagome lattice
in invariant under $a\rightarrow b\rightarrow c\rightarrow a$, but
the wavefunction acquires a negative sign under a C$_6$ rotation.
Thus the fermionic ground state possesses an $f$-wave symmetry.
This property is also obeyed by the localized state discussed in
Ref. \onlinecite{Bergman2008}. Since this property is maintained
by any $\vk$-state, it suggests that any fermionic state with a
filling fraction $\nu<1/3$ also has this character.

\subsection{Real space wavefunctions}
The Bloch solution, allows us to write down the solution in real
space as\bea\label{eq:RS}
\psi^a_{\vk}(\vrr)&=&\hat k_3\phi_{\vk}(\vrr),\nonumber\\
\psi^b_{\vk}(\vrr)&=&-\hat k_1\phi_{\vk}(\vrr),\nonumber\\
\psi^c_{\vk}(\vrr)&=&\hat k_2\phi_{\vk}(\vrr), \eea where $\hat
k_i\equiv-i\mb a_i\cdot\partial_{\vrr}$, and $\phi_{\vk}(\vrr)$ is
a function that solves the characteristic equation of the Kagome
Hamiltonian. Because of the structure of Eq. (\ref{eq:char2s}), we
see that $\phi_{\vk}(\vrr)=e^{i\vk\cdot\vrr}$.

\section{The Chern-Simons flux and the covariant momentum}
As introduced in the main text, our MFA introduces two fluxes
$\phi_C$ and $\phi_T$ (which are eventually set equal). While
$\phi_C$ is simply imposed onto the model, $\phi_T$ originates
from a vector potential $\mb A(\vrr)$ and grows with area
(Maxwell-type). We remind the reader that  $\mb A(\vrr)$ is the
one that is to be used in creating the covariant momentum
$\vp=-i\partial_{\vrr}+\mb A(\vrr)$. The corresponding
translation operators have the following properties:
\bea\label{eq:def} p_1&=&p_2+p_3,\nonumber\\
e^{-ip_2}e^{-ip_3}e^{ip_1}&=&e^{-i\phi_T}~=~e^{-ip_1}e^{ip_2}e^{ip_3},\nonumber\\
e^{-ip_3}e^{-ip_2}e^{ip_1}&=&e^{i\phi_T}~=~e^{-ip_1}e^{ip_3}e^{ip_2},\nonumber\\
e^{-ip_3}e^{ip_1}e^{ip_3}e^{-ip_1}&=& e^{-2i\phi_T},\nonumber\\
e^{-2ip_3}e^{2ip_1}e^{2ip_3}e^{-2ip_1}&=&
e^{-8i\phi_T}~(\text{area of unit cell}),\nonumber\\
e^{ip_3}e^{ip_1}e^{ip_2}&=& e^{3i\phi_T}e^{2ip_1},\nonumber\\
e^{ip_2}e^{ip_1}e^{ip_3}&=& e^{-3i\phi_T}e^{2ip_1},\nonumber\\
e^{ip_3}e^{ip_1}e^{ip_2}e^{-ip_3}e^{-ip_1}e^{-ip_2}&=&e^{6i\phi_T}~
(\text{area of the hexagon}).\nonumber \eea Here $p_i=\vp\cdot\mb
a_i$. We further have the following commutation relations for
$i,j\in\{1,2,3\}$: \bea\label{eq:commie}
\left[k_i,k_j\right]&=&0,\nonumber\\
\left[p_1,p_2\right]&=&-2i\phi_T,\nonumber\\
\left[p_2,p_3\right]&=&2i\phi_T,\nonumber\\
\left[p_3,p_1\right]&=&-2i\phi_T,\nonumber\\
\eea

The flux $\phi_C$ is introduced to account for the internal
modulation and is incorporated directly in the Hamiltonian as
shown in Eq (9) of the main text. This is necessary because the
continuum limit is obtained form the Bloch solution. The flux
$\phi_C$, a property of the unit cell itself, cannot be accounted
for by introducing a position dependent gauge field like $\mb
A(\vrr)$.

The translational operator on a lattice taking a fermion from
$\vrr_1$ to $\vrr_2$ is $T_{\vrr_2\vrr_1}\equiv
c^{\dag}_{\vrr_2}c_{\vrr_1}$. For a triangular loop
$L_T:~a\rightarrow b\rightarrow c$, it follows that $T(L_T)\equiv
T_{\vrr_a\vrr_c}T_{\vrr_c\vrr_b}T_{\vrr_b\vrr_a}$ is the same as
$T(L_T')\equiv T_{\vrr_a\vrr_b}T_{\vrr_b\vrr_c}T_{\vrr_c\vrr_a}$.
If we couple the fermions to a Maxwell-type gauge field where the
flux grows with the area, then $T_{\vrr_2\vrr_1}\rightarrow
e^{i\Lambda^M_{\vrr_2\vrr_1}}c^{\dag}_{\vrr_2}c_{\vrr_1}$
and\bea\label{eq:5}
T^M(L_T)&=&e^{i\phi_T}T(L_T)\nonumber\\
T^{M}(L_T')&=&e^{-i\phi_T}T(L_T)\nonumber\\
\phi_T&=& \Lambda^M_{\vrr_a\vrr_c}+\Lambda^M_{\vrr_c\vrr_b}
+\Lambda^M_{\vrr_b\vrr_a}=\int\mb{A}\cdot d\mb{l}.\eea Similarly,
we may consider the hexagonal loop which yields
$T^M(L_H)=e^{i6\phi_T}T(L_H)$.

On implementing the CS flux as shown in Figs. 2c and 3 of the main
text, we see that \bea
T^{CS}(L_T)&=&e^{i(\phi_T-\phi_C)}T(L_T)=T(L_T),\nonumber\\
T^{CS}(L_H)&=&e^{i(6\phi_T+2\phi_C)}T(L_T)=e^{8i\phi_T}T(L_T).\nonumber\\
\eea The last equality is obtained by setting $\phi_T=\phi_C$.
Equating the total flux through the unit cell $8\phi_T$ to
$6\pi\nu$, we arrive at the relation
\beq\label{eq:flux}\phi_T=\phi_C=3\pi\nu/4.\eeq

\section{The flat band wavefunction in a Kagome lattice with CS flux}
Before deriving the case with the CS flux, we explicitly derive
Eq. (\ref{eq:RS}). This is informative and the derivation with the
CS flux follows similar lines. Plugging the flat band eigenvalue
to the Hamiltonian we see that the flat band wavefunction
components $\psi^{a}$ satisfy \bea\label{eq:supp1}
k_2(k_2\psi^a-k_3\psi^c)&=&0,\nonumber\\
k_3(k_3\psi^b+k_1\psi^a)&=&0,\nonumber\\
k_1(k_1\psi^c+k_2\psi^b)&=&0. \eea It is useful note that other
equations can be generated using $a\rightarrow b\rightarrow c
\rightarrow a $; $\{H_1,H_1^*\}\rightarrow \{H_2^*,H_2\}$,
$\{H_2,H_2^*\}\rightarrow\{H_3,H_3^*\}$, $\{H_3,H_3^*\}\rightarrow
\{H_1^*,H_1\}$; and $k_1\rightarrow k_2$, $k_2 \rightarrow -k_3$,
$k_3 \rightarrow k_1$. This implies that
$k_1(\psi^a+\psi^b+\psi^c)=$const. Since $\vk\rightarrow
-i\partial_{\vrr}$, normalizability over the whole space not only
requires const$=0$, but $\psi^a+\psi^b+\psi^c=0$. The only
combination that satisfies the Hamiltonian is then given by Eq.
(\ref{eq:RS}).

It is sometimes inconvenient to have the components of the
wavefunctions expressed as operators. To remedy this the action of
the operator $\hat R_{\vrr}$ can be implemented by convoluting
with the Greens' function of the operator $R(\vrr-\vrr')$. Thus
\beq\label{eq:GF} \hat R_{\vrr}
f(\vrr)=\int_{\vrr'}R(\vrr-\vrr')f(\vrr'). \eeq If $\hat
R_{\vrr}=-i\partial_{\vrr}$, then
$R(\vrr-\vrr')=-i\partial_{\vrr}\delta(\vrr-\vrr')=i\partial_{\vrr'}\delta(\vrr-\vrr')$.

When a similar analysis is carried out for $\hat H^{\rm MFA}$ with
the CS flux attached, we end up with Eq. (\ref{eq:supp1}) but with
$\vk\rightarrow\vp\equiv -i\partial_{\vrr}+\mb A(\vrr)$ (only
for $\phi_T=\phi_C$). The cyclic interchange also works the same
way with $\vk\rightarrow\vp$, and with an addition of
$\phi_C\rightarrow -\phi_C$. Just like before, normalizability
will enforce that $\psi^a+\psi^b+\psi^c=0$. This result is
independent of the choice of gauge for writing down $\mb
A(\vrr)$. Thus the flat band wavefunction can be written as
\bea\label{eq:RS2}
\psi^a_{\vk}(\vrr)&=&\hat p_3\phi_{i}(\vrr),\nonumber\\
\psi^b_{\vk}(\vrr)&=&-\hat p_1\phi_{i}(\vrr),\nonumber\\
\psi^c_{\vk}(\vrr)&=&\hat p_2\phi_{i}(\vrr), \eea where $i$ is
some index denoting the quantum state (which is no longer the
momentum). It is worth noting that when this wavefunction is
substituted back into the Hamiltonian equations for $E=-2J_1$, we
get\bea\label{eq:supp2}
0&=&2\psi^a+H_2e^{i\phi_C}\psi^b+H_1\psi^c + \mathcal{O}(p^3)\nonumber\\
&=&2p_3 + (2-p_2^2)(1+i\phi_C)(-p_1)+(2-p_1^2)p_2 + \mathcal{O}(p^3)\nonumber\\
&=&2(p_3+p_2-p_1)-2ip_1\phi_C + p_2^2p_1-p_1^2p_2 + \mathcal{O}(p^3)\nonumber\\
&=&-2ip_1(\phi_C-\phi_T) +\mathcal{O}(p^3)\nonumber\\
&=&0+\mathcal{O}(p^3). \eea Note that since our equations are
derived correct to $\mathcal{O}(p^2)$, we conclude that the
wavefunction guess in Eq. (\ref{eq:RS2}) is correct to
$\mathcal{O}(p^2)$.

\section{Normalization of the flat band wavefunction}
We require $\int_{\vrr}\Psi^{\dag}(\vrr)\Psi(\vrr) =1$. From Eq. (12) of the main text and using the form of $f(\vrr)$, we see that \bea\label{eq:app1}\mathcal{N}^2&=&\int_{\vrr}\left[\sum_i \{p_i^*f(\vrr)\}\{p_if(\vrr)\}\right]\nonumber\\
&=&\int_{\vrr}\left[\sum_i \{k_if(\vrr)\}^2 + \{A_if(\vrr)\}^2\right]\nonumber\\
&=&\frac{3}{2}\int_{\vrr}\left[\{\partial_xf(\vrr)\}^2 + \{\partial_yf(\vrr)\}^2+A^2f^2(\vrr)\right]\nonumber\\
&=&\frac34\left[1+gB^2l^4_{CS}\right], \eea where $g$ is a gauge
dependent constant factor. If $\mb A$ is chosen in Landau gauge,
$g=1/2$. If $\mb A$ is chosen in symmetric gauge, $g=1/4$.

\begin{bibliography}{}

\end{bibliography}


\begin{thebibliography}{53}
\expandafter\ifx\csname natexlab\endcsname\relax\def\natexlab#1{#1}\fi
\expandafter\ifx\csname bibnamefont\endcsname\relax
  \def\bibnamefont#1{#1}\fi
\expandafter\ifx\csname bibfnamefont\endcsname\relax
  \def\bibfnamefont#1{#1}\fi
\expandafter\ifx\csname citenamefont\endcsname\relax
  \def\citenamefont#1{#1}\fi
\expandafter\ifx\csname url\endcsname\relax
  \def\url#1{\texttt{#1}}\fi
\expandafter\ifx\csname urlprefix\endcsname\relax\def\urlprefix{URL }\fi
\providecommand{\bibinfo}[2]{#2}
\providecommand{\eprint}[2][]{\url{#2}}

\bibitem[{\citenamefont{Han et~al.}(2012)\citenamefont{Han, Helton, Chu,
  Nocera, Rodriguez-Rivera, Broholm, and Lee}}]{Han2012}
\bibinfo{author}{\bibfnamefont{T.-H.} \bibnamefont{Han}},
  \bibinfo{author}{\bibfnamefont{J.~S.} \bibnamefont{Helton}},
  \bibinfo{author}{\bibfnamefont{S.}~\bibnamefont{Chu}},
  \bibinfo{author}{\bibfnamefont{D.~G.} \bibnamefont{Nocera}},
  \bibinfo{author}{\bibfnamefont{J.~A.} \bibnamefont{Rodriguez-Rivera}},
  \bibinfo{author}{\bibfnamefont{C.}~\bibnamefont{Broholm}}, \bibnamefont{and}
  \bibinfo{author}{\bibfnamefont{Y.~S.} \bibnamefont{Lee}},
  \bibinfo{journal}{Nature} \textbf{\bibinfo{volume}{492}}, \bibinfo{pages}{406
  EP } (\bibinfo{year}{2012}),
  \urlprefix\url{http://dx.doi.org/10.1038/nature11659}.

\bibitem[{\citenamefont{Punk et~al.}(2014)\citenamefont{Punk, Chowdhury, and
  Sachdev}}]{Punk2014}
\bibinfo{author}{\bibfnamefont{M.}~\bibnamefont{Punk}},
  \bibinfo{author}{\bibfnamefont{D.}~\bibnamefont{Chowdhury}},
  \bibnamefont{and} \bibinfo{author}{\bibfnamefont{S.}~\bibnamefont{Sachdev}},
  \bibinfo{journal}{Nature Physics} \textbf{\bibinfo{volume}{10}},
  \bibinfo{pages}{289 EP } (\bibinfo{year}{2014}),
  \urlprefix\url{http://dx.doi.org/10.1038/nphys2887}.

\bibitem[{\citenamefont{Ran et~al.}(2007)\citenamefont{Ran, Hermele, Lee, and
  Wen}}]{Ran2007}
\bibinfo{author}{\bibfnamefont{Y.}~\bibnamefont{Ran}},
  \bibinfo{author}{\bibfnamefont{M.}~\bibnamefont{Hermele}},
  \bibinfo{author}{\bibfnamefont{P.~A.} \bibnamefont{Lee}}, \bibnamefont{and}
  \bibinfo{author}{\bibfnamefont{X.-G.} \bibnamefont{Wen}},
  \bibinfo{journal}{Phys. Rev. Lett.} \textbf{\bibinfo{volume}{98}},
  \bibinfo{pages}{117205} (\bibinfo{year}{2007}),
  \urlprefix\url{https://link.aps.org/doi/10.1103/PhysRevLett.98.117205}.

\bibitem[{\citenamefont{Iqbal et~al.}(2011)\citenamefont{Iqbal, Becca, and
  Poilblanc}}]{Iqbal2011}
\bibinfo{author}{\bibfnamefont{Y.}~\bibnamefont{Iqbal}},
  \bibinfo{author}{\bibfnamefont{F.}~\bibnamefont{Becca}}, \bibnamefont{and}
  \bibinfo{author}{\bibfnamefont{D.}~\bibnamefont{Poilblanc}},
  \bibinfo{journal}{Phys. Rev. B} \textbf{\bibinfo{volume}{84}},
  \bibinfo{pages}{020407} (\bibinfo{year}{2011}),
  \urlprefix\url{https://link.aps.org/doi/10.1103/PhysRevB.84.020407}.

\bibitem[{\citenamefont{Yan et~al.}(2011)\citenamefont{Yan, Huse, and
  White}}]{Yan2011}
\bibinfo{author}{\bibfnamefont{S.}~\bibnamefont{Yan}},
  \bibinfo{author}{\bibfnamefont{D.~A.} \bibnamefont{Huse}}, \bibnamefont{and}
  \bibinfo{author}{\bibfnamefont{S.~R.} \bibnamefont{White}},
  \bibinfo{journal}{Science} \textbf{\bibinfo{volume}{332}},
  \bibinfo{pages}{1173} (\bibinfo{year}{2011}), ISSN \bibinfo{issn}{0036-8075},
  \urlprefix\url{http://science.sciencemag.org/content/332/6034/1173}.

\bibitem[{\citenamefont{Yang et~al.}(1993)\citenamefont{Yang, Warman, and
  Girvin}}]{Yang1993}
\bibinfo{author}{\bibfnamefont{K.}~\bibnamefont{Yang}},
  \bibinfo{author}{\bibfnamefont{L.~K.} \bibnamefont{Warman}},
  \bibnamefont{and} \bibinfo{author}{\bibfnamefont{S.~M.}
  \bibnamefont{Girvin}}, \bibinfo{journal}{Phys. Rev. Lett.}
  \textbf{\bibinfo{volume}{70}}, \bibinfo{pages}{2641} (\bibinfo{year}{1993}),
  \urlprefix\url{https://link.aps.org/doi/10.1103/PhysRevLett.70.2641}.

\bibitem[{\citenamefont{Messio et~al.}(2012)\citenamefont{Messio, Bernu, and
  Lhuillier}}]{Messio2012}
\bibinfo{author}{\bibfnamefont{L.}~\bibnamefont{Messio}},
  \bibinfo{author}{\bibfnamefont{B.}~\bibnamefont{Bernu}}, \bibnamefont{and}
  \bibinfo{author}{\bibfnamefont{C.}~\bibnamefont{Lhuillier}},
  \bibinfo{journal}{Phys. Rev. Lett.} \textbf{\bibinfo{volume}{108}},
  \bibinfo{pages}{207204} (\bibinfo{year}{2012}),
  \urlprefix\url{https://link.aps.org/doi/10.1103/PhysRevLett.108.207204}.

\bibitem[{\citenamefont{He et~al.}(2014)\citenamefont{He, Sheng, and
  Chen}}]{He2014}
\bibinfo{author}{\bibfnamefont{Y.-C.} \bibnamefont{He}},
  \bibinfo{author}{\bibfnamefont{D.~N.} \bibnamefont{Sheng}}, \bibnamefont{and}
  \bibinfo{author}{\bibfnamefont{Y.}~\bibnamefont{Chen}},
  \bibinfo{journal}{Phys. Rev. Lett.} \textbf{\bibinfo{volume}{112}},
  \bibinfo{pages}{137202} (\bibinfo{year}{2014}),
  \urlprefix\url{https://link.aps.org/doi/10.1103/PhysRevLett.112.137202}.

\bibitem[{\citenamefont{Gong et~al.}(2014)\citenamefont{Gong, Zhu, and
  Sheng}}]{Gong2014}
\bibinfo{author}{\bibfnamefont{S.-S.} \bibnamefont{Gong}},
  \bibinfo{author}{\bibfnamefont{W.}~\bibnamefont{Zhu}}, \bibnamefont{and}
  \bibinfo{author}{\bibfnamefont{D.~N.} \bibnamefont{Sheng}},
  \bibinfo{journal}{Scientific Reports} \textbf{\bibinfo{volume}{4}},
  \bibinfo{pages}{6317 EP } (\bibinfo{year}{2014}), \bibinfo{note}{article},
  \urlprefix\url{http://dx.doi.org/10.1038/srep06317}.

\bibitem[{\citenamefont{Bauer et~al.}(2014)\citenamefont{Bauer, Cincio, Keller,
  Dolfi, Vidal, Trebst, and Ludwig}}]{Bauer2014}
\bibinfo{author}{\bibfnamefont{B.}~\bibnamefont{Bauer}},
  \bibinfo{author}{\bibfnamefont{L.}~\bibnamefont{Cincio}},
  \bibinfo{author}{\bibfnamefont{B.~P.} \bibnamefont{Keller}},
  \bibinfo{author}{\bibfnamefont{M.}~\bibnamefont{Dolfi}},
  \bibinfo{author}{\bibfnamefont{G.}~\bibnamefont{Vidal}},
  \bibinfo{author}{\bibfnamefont{S.}~\bibnamefont{Trebst}}, \bibnamefont{and}
  \bibinfo{author}{\bibfnamefont{A.~W.~W.} \bibnamefont{Ludwig}},
  \bibinfo{journal}{Nature Communications} \textbf{\bibinfo{volume}{5}},
  \bibinfo{pages}{5137 EP } (\bibinfo{year}{2014}), \bibinfo{note}{article},
  \urlprefix\url{http://dx.doi.org/10.1038/ncomms6137}.

\bibitem[{\citenamefont{Kumar et~al.}(2014)\citenamefont{Kumar, Sun, and
  Fradkin}}]{Kumar2014}
\bibinfo{author}{\bibfnamefont{K.}~\bibnamefont{Kumar}},
  \bibinfo{author}{\bibfnamefont{K.}~\bibnamefont{Sun}}, \bibnamefont{and}
  \bibinfo{author}{\bibfnamefont{E.}~\bibnamefont{Fradkin}},
  \bibinfo{journal}{Phys. Rev. B} \textbf{\bibinfo{volume}{90}},
  \bibinfo{pages}{174409} (\bibinfo{year}{2014}),
  \urlprefix\url{https://link.aps.org/doi/10.1103/PhysRevB.90.174409}.

\bibitem[{\citenamefont{Kumar et~al.}(2015)\citenamefont{Kumar, Sun, and
  Fradkin}}]{Kumar2015}
\bibinfo{author}{\bibfnamefont{K.}~\bibnamefont{Kumar}},
  \bibinfo{author}{\bibfnamefont{K.}~\bibnamefont{Sun}}, \bibnamefont{and}
  \bibinfo{author}{\bibfnamefont{E.}~\bibnamefont{Fradkin}},
  \bibinfo{journal}{Phys. Rev. B} \textbf{\bibinfo{volume}{92}},
  \bibinfo{pages}{094433} (\bibinfo{year}{2015}),
  \urlprefix\url{https://link.aps.org/doi/10.1103/PhysRevB.92.094433}.

\bibitem[{\citenamefont{Zaletel et~al.}(2016)\citenamefont{Zaletel, Zhu, Lu,
  Vishwanath, and White}}]{Zalatel2016}
\bibinfo{author}{\bibfnamefont{M.~P.} \bibnamefont{Zaletel}},
  \bibinfo{author}{\bibfnamefont{Z.}~\bibnamefont{Zhu}},
  \bibinfo{author}{\bibfnamefont{Y.-M.} \bibnamefont{Lu}},
  \bibinfo{author}{\bibfnamefont{A.}~\bibnamefont{Vishwanath}},
  \bibnamefont{and} \bibinfo{author}{\bibfnamefont{S.~R.} \bibnamefont{White}},
  \bibinfo{journal}{Phys. Rev. Lett.} \textbf{\bibinfo{volume}{116}},
  \bibinfo{pages}{197203} (\bibinfo{year}{2016}),
  \urlprefix\url{https://link.aps.org/doi/10.1103/PhysRevLett.116.197203}.

\bibitem[{\citenamefont{Pereira and Bieri}(2018)}]{Pereira2018}
\bibinfo{author}{\bibfnamefont{R.~G.} \bibnamefont{Pereira}} \bibnamefont{and}
  \bibinfo{author}{\bibfnamefont{S.}~\bibnamefont{Bieri}},
  \bibinfo{journal}{SciPost Phys.} \textbf{\bibinfo{volume}{4}},
  \bibinfo{pages}{004} (\bibinfo{year}{2018}),
  \urlprefix\url{https://scipost.org/10.21468/SciPostPhys.4.1.004}.

\bibitem[{\citenamefont{Kalmeyer and Laughlin}(1987)}]{Kalmeyer1987}
\bibinfo{author}{\bibfnamefont{V.}~\bibnamefont{Kalmeyer}} \bibnamefont{and}
  \bibinfo{author}{\bibfnamefont{R.~B.} \bibnamefont{Laughlin}},
  \bibinfo{journal}{Phys. Rev. Lett.} \textbf{\bibinfo{volume}{59}},
  \bibinfo{pages}{2095} (\bibinfo{year}{1987}),
  \urlprefix\url{https://link.aps.org/doi/10.1103/PhysRevLett.59.2095}.

\bibitem[{\citenamefont{Wen et~al.}(1989)\citenamefont{Wen, Wilczek, and
  Zee}}]{Wen1989}
\bibinfo{author}{\bibfnamefont{X.~G.} \bibnamefont{Wen}},
  \bibinfo{author}{\bibfnamefont{F.}~\bibnamefont{Wilczek}}, \bibnamefont{and}
  \bibinfo{author}{\bibfnamefont{A.}~\bibnamefont{Zee}},
  \bibinfo{journal}{Phys. Rev. B} \textbf{\bibinfo{volume}{39}},
  \bibinfo{pages}{11413} (\bibinfo{year}{1989}),
  \urlprefix\url{https://link.aps.org/doi/10.1103/PhysRevB.39.11413}.

\bibitem[{\citenamefont{Khveshchenko and Wiegmann}(1989)}]{KHVESHCHENKO1989}
\bibinfo{author}{\bibfnamefont{D.~V.} \bibnamefont{Khveshchenko}}
  \bibnamefont{and} \bibinfo{author}{\bibfnamefont{P.~B.}
  \bibnamefont{Wiegmann}}, \bibinfo{journal}{Modern Physics Letters B}
  \textbf{\bibinfo{volume}{03}}, \bibinfo{pages}{1383} (\bibinfo{year}{1989}),
  \urlprefix\url{https://doi.org/10.1142/S0217984989002089}.

\bibitem[{\citenamefont{Khveshchenko and Wiegmann}(1990)}]{KHVESHCHENKO1990}
\bibinfo{author}{\bibfnamefont{D.~V.} \bibnamefont{Khveshchenko}}
  \bibnamefont{and} \bibinfo{author}{\bibfnamefont{P.~B.}
  \bibnamefont{Wiegmann}}, \bibinfo{journal}{Modern Physics Letters B}
  \textbf{\bibinfo{volume}{04}}, \bibinfo{pages}{17} (\bibinfo{year}{1990}),
  \urlprefix\url{https://doi.org/10.1142/S0217984990000040}.

\bibitem[{\citenamefont{Kitaev}(2006)}]{KITAEV2006}
\bibinfo{author}{\bibfnamefont{A.}~\bibnamefont{Kitaev}},
  \bibinfo{journal}{Annals of Physics} \textbf{\bibinfo{volume}{321}},
  \bibinfo{pages}{2 } (\bibinfo{year}{2006}), ISSN \bibinfo{issn}{0003-4916},
  \bibinfo{note}{january Special Issue},
  \urlprefix\url{http://www.sciencedirect.com/science/article/pii/S0003491605002381}.

\bibitem[{\citenamefont{Yao and Kivelson}(2007)}]{Yao2007}
\bibinfo{author}{\bibfnamefont{H.}~\bibnamefont{Yao}} \bibnamefont{and}
  \bibinfo{author}{\bibfnamefont{S.~A.} \bibnamefont{Kivelson}},
  \bibinfo{journal}{Phys. Rev. Lett.} \textbf{\bibinfo{volume}{99}},
  \bibinfo{pages}{247203} (\bibinfo{year}{2007}),
  \urlprefix\url{https://link.aps.org/doi/10.1103/PhysRevLett.99.247203}.

\bibitem[{\citenamefont{Xu and Sachdev}(2009)}]{Xu2009}
\bibinfo{author}{\bibfnamefont{C.}~\bibnamefont{Xu}} \bibnamefont{and}
  \bibinfo{author}{\bibfnamefont{S.}~\bibnamefont{Sachdev}},
  \bibinfo{journal}{Phys. Rev. B} \textbf{\bibinfo{volume}{79}},
  \bibinfo{pages}{064405} (\bibinfo{year}{2009}),
  \urlprefix\url{https://link.aps.org/doi/10.1103/PhysRevB.79.064405}.

\bibitem[{\citenamefont{Wietek and L\"auchli}(2017)}]{Wietek2017}
\bibinfo{author}{\bibfnamefont{A.}~\bibnamefont{Wietek}} \bibnamefont{and}
  \bibinfo{author}{\bibfnamefont{A.~M.} \bibnamefont{L\"auchli}},
  \bibinfo{journal}{Phys. Rev. B} \textbf{\bibinfo{volume}{95}},
  \bibinfo{pages}{035141} (\bibinfo{year}{2017}),
  \urlprefix\url{https://link.aps.org/doi/10.1103/PhysRevB.95.035141}.

\bibitem[{\citenamefont{Werman et~al.}(2018)\citenamefont{Werman, Chatterjee,
  Morampudi, and Berg}}]{Werman2018}
\bibinfo{author}{\bibfnamefont{Y.}~\bibnamefont{Werman}},
  \bibinfo{author}{\bibfnamefont{S.}~\bibnamefont{Chatterjee}},
  \bibinfo{author}{\bibfnamefont{S.~C.} \bibnamefont{Morampudi}},
  \bibnamefont{and} \bibinfo{author}{\bibfnamefont{E.}~\bibnamefont{Berg}},
  \bibinfo{journal}{Phys. Rev. X} \textbf{\bibinfo{volume}{8}},
  \bibinfo{pages}{031064} (\bibinfo{year}{2018}),
  \urlprefix\url{https://link.aps.org/doi/10.1103/PhysRevX.8.031064}.

\bibitem[{\citenamefont{Balents}(2010)}]{Balents2010}
\bibinfo{author}{\bibfnamefont{L.}~\bibnamefont{Balents}},
  \bibinfo{journal}{Nature} \textbf{\bibinfo{volume}{464}}, \bibinfo{pages}{199
  EP } (\bibinfo{year}{2010}),
  \urlprefix\url{http://dx.doi.org/10.1038/nature08917}.

\bibitem[{\citenamefont{Savary and Balents}(2017)}]{Savary2017}
\bibinfo{author}{\bibfnamefont{L.}~\bibnamefont{Savary}} \bibnamefont{and}
  \bibinfo{author}{\bibfnamefont{L.}~\bibnamefont{Balents}},
  \bibinfo{journal}{Reports on Progress in Physics}
  \textbf{\bibinfo{volume}{80}}, \bibinfo{pages}{016502}
  (\bibinfo{year}{2017}),
  \urlprefix\url{http://stacks.iop.org/0034-4885/80/i=1/a=016502}.

\bibitem[{\citenamefont{Zhou et~al.}(2017)\citenamefont{Zhou, Kanoda, and
  Ng}}]{Zhou2017}
\bibinfo{author}{\bibfnamefont{Y.}~\bibnamefont{Zhou}},
  \bibinfo{author}{\bibfnamefont{K.}~\bibnamefont{Kanoda}}, \bibnamefont{and}
  \bibinfo{author}{\bibfnamefont{T.-K.} \bibnamefont{Ng}},
  \bibinfo{journal}{Rev. Mod. Phys.} \textbf{\bibinfo{volume}{89}},
  \bibinfo{pages}{025003} (\bibinfo{year}{2017}),
  \urlprefix\url{https://link.aps.org/doi/10.1103/RevModPhys.89.025003}.

\bibitem[{\citenamefont{Benton et~al.}(2018)\citenamefont{Benton, Jaubert,
  Singh, Oitmaa, and Shannon}}]{Benton2018}
\bibinfo{author}{\bibfnamefont{O.}~\bibnamefont{Benton}},
  \bibinfo{author}{\bibfnamefont{L.~D.~C.} \bibnamefont{Jaubert}},
  \bibinfo{author}{\bibfnamefont{R.~R.~P.} \bibnamefont{Singh}},
  \bibinfo{author}{\bibfnamefont{J.}~\bibnamefont{Oitmaa}}, \bibnamefont{and}
  \bibinfo{author}{\bibfnamefont{N.}~\bibnamefont{Shannon}},
  \bibinfo{journal}{Phys. Rev. Lett.} \textbf{\bibinfo{volume}{121}},
  \bibinfo{pages}{067201} (\bibinfo{year}{2018}),
  \urlprefix\url{https://link.aps.org/doi/10.1103/PhysRevLett.121.067201}.

\bibitem[{\citenamefont{Norman}(2016)}]{Norman2016}
\bibinfo{author}{\bibfnamefont{M.~R.} \bibnamefont{Norman}},
  \bibinfo{journal}{Rev. Mod. Phys.} \textbf{\bibinfo{volume}{88}},
  \bibinfo{pages}{041002} (\bibinfo{year}{2016}),
  \urlprefix\url{https://link.aps.org/doi/10.1103/RevModPhys.88.041002}.

\bibitem[{\citenamefont{Pohle et~al.}(2017)\citenamefont{Pohle, Yan, and
  Shannon}}]{Pohle2017}
\bibinfo{author}{\bibfnamefont{R.}~\bibnamefont{Pohle}},
  \bibinfo{author}{\bibfnamefont{H.}~\bibnamefont{Yan}}, \bibnamefont{and}
  \bibinfo{author}{\bibfnamefont{N.}~\bibnamefont{Shannon}},
  \bibinfo{journal}{arXiv}  (\bibinfo{year}{2017}),
  \urlprefix\url{https://arxiv.org/abs/1711.03778}.

\bibitem[{\citenamefont{Bergman et~al.}(2008)\citenamefont{Bergman, Wu, and
  Balents}}]{Bergman2008}
\bibinfo{author}{\bibfnamefont{D.~L.} \bibnamefont{Bergman}},
  \bibinfo{author}{\bibfnamefont{C.}~\bibnamefont{Wu}}, \bibnamefont{and}
  \bibinfo{author}{\bibfnamefont{L.}~\bibnamefont{Balents}},
  \bibinfo{journal}{Phys. Rev. B} \textbf{\bibinfo{volume}{78}},
  \bibinfo{pages}{125104} (\bibinfo{year}{2008}),
  \urlprefix\url{https://link.aps.org/doi/10.1103/PhysRevB.78.125104}.

\bibitem[{\citenamefont{Huber and Altman}(2010)}]{Huber2010}
\bibinfo{author}{\bibfnamefont{S.~D.} \bibnamefont{Huber}} \bibnamefont{and}
  \bibinfo{author}{\bibfnamefont{E.}~\bibnamefont{Altman}},
  \bibinfo{journal}{Phys. Rev. B} \textbf{\bibinfo{volume}{82}},
  \bibinfo{pages}{184502} (\bibinfo{year}{2010}),
  \urlprefix\url{https://link.aps.org/doi/10.1103/PhysRevB.82.184502}.

\bibitem[{\citenamefont{Jain}(1989)}]{Jain1989}
\bibinfo{author}{\bibfnamefont{J.~K.} \bibnamefont{Jain}},
  \bibinfo{journal}{Phys. Rev. Lett.} \textbf{\bibinfo{volume}{63}},
  \bibinfo{pages}{199} (\bibinfo{year}{1989}),
  \urlprefix\url{https://link.aps.org/doi/10.1103/PhysRevLett.63.199}.

\bibitem[{\citenamefont{Lopez and Fradkin}(1991)}]{Lopez1991}
\bibinfo{author}{\bibfnamefont{A.}~\bibnamefont{Lopez}} \bibnamefont{and}
  \bibinfo{author}{\bibfnamefont{E.}~\bibnamefont{Fradkin}},
  \bibinfo{journal}{Phys. Rev. B} \textbf{\bibinfo{volume}{44}},
  \bibinfo{pages}{5246} (\bibinfo{year}{1991}),
  \urlprefix\url{https://link.aps.org/doi/10.1103/PhysRevB.44.5246}.

\bibitem[{\citenamefont{Halperin et~al.}(1993)\citenamefont{Halperin, Lee, and
  Read}}]{Halperin1993}
\bibinfo{author}{\bibfnamefont{B.~I.} \bibnamefont{Halperin}},
  \bibinfo{author}{\bibfnamefont{P.~A.} \bibnamefont{Lee}}, \bibnamefont{and}
  \bibinfo{author}{\bibfnamefont{N.}~\bibnamefont{Read}},
  \bibinfo{journal}{Phys. Rev. B} \textbf{\bibinfo{volume}{47}},
  \bibinfo{pages}{7312} (\bibinfo{year}{1993}),
  \urlprefix\url{https://link.aps.org/doi/10.1103/PhysRevB.47.7312}.

\bibitem[{\citenamefont{Zhang et~al.}(1989)\citenamefont{Zhang, Hansson, and
  Kivelson}}]{PhysRevLett.62.82}
\bibinfo{author}{\bibfnamefont{S.~C.} \bibnamefont{Zhang}},
  \bibinfo{author}{\bibfnamefont{T.~H.} \bibnamefont{Hansson}},
  \bibnamefont{and} \bibinfo{author}{\bibfnamefont{S.}~\bibnamefont{Kivelson}},
  \bibinfo{journal}{Phys. Rev. Lett.} \textbf{\bibinfo{volume}{62}},
  \bibinfo{pages}{82} (\bibinfo{year}{1989}),
  \urlprefix\url{https://link.aps.org/doi/10.1103/PhysRevLett.62.82}.

\bibitem[{\citenamefont{Sedrakyan et~al.}(2012)\citenamefont{Sedrakyan,
  Kamenev, and Glazman}}]{Sedrakyan2012}
\bibinfo{author}{\bibfnamefont{T.~A.} \bibnamefont{Sedrakyan}},
  \bibinfo{author}{\bibfnamefont{A.}~\bibnamefont{Kamenev}}, \bibnamefont{and}
  \bibinfo{author}{\bibfnamefont{L.~I.} \bibnamefont{Glazman}},
  \bibinfo{journal}{Phys. Rev. A} \textbf{\bibinfo{volume}{86}},
  \bibinfo{pages}{063639} (\bibinfo{year}{2012}),
  \urlprefix\url{https://link.aps.org/doi/10.1103/PhysRevA.86.063639}.

\bibitem[{\citenamefont{Sedrakyan
  et~al.}(2015{\natexlab{a}})\citenamefont{Sedrakyan, Galitski, and
  Kamenev}}]{Sedrakyan2015_2}
\bibinfo{author}{\bibfnamefont{T.~A.} \bibnamefont{Sedrakyan}},
  \bibinfo{author}{\bibfnamefont{V.~M.} \bibnamefont{Galitski}},
  \bibnamefont{and} \bibinfo{author}{\bibfnamefont{A.}~\bibnamefont{Kamenev}},
  \bibinfo{journal}{Phys. Rev. Lett.} \textbf{\bibinfo{volume}{115}},
  \bibinfo{pages}{195301} (\bibinfo{year}{2015}{\natexlab{a}}),
  \urlprefix\url{https://link.aps.org/doi/10.1103/PhysRevLett.115.195301}.

\bibitem[{\citenamefont{Sedrakyan et~al.}(2014)\citenamefont{Sedrakyan,
  Glazman, and Kamenev}}]{Sedrakyan2014}
\bibinfo{author}{\bibfnamefont{T.~A.} \bibnamefont{Sedrakyan}},
  \bibinfo{author}{\bibfnamefont{L.~I.} \bibnamefont{Glazman}},
  \bibnamefont{and} \bibinfo{author}{\bibfnamefont{A.}~\bibnamefont{Kamenev}},
  \bibinfo{journal}{Phys. Rev. B} \textbf{\bibinfo{volume}{89}},
  \bibinfo{pages}{201112} (\bibinfo{year}{2014}),
  \urlprefix\url{https://link.aps.org/doi/10.1103/PhysRevB.89.201112}.

\bibitem[{\citenamefont{Sedrakyan et~al.}(2017)\citenamefont{Sedrakyan,
  Galitski, and Kamenev}}]{Sedrakyan2017}
\bibinfo{author}{\bibfnamefont{T.~A.} \bibnamefont{Sedrakyan}},
  \bibinfo{author}{\bibfnamefont{V.~M.} \bibnamefont{Galitski}},
  \bibnamefont{and} \bibinfo{author}{\bibfnamefont{A.}~\bibnamefont{Kamenev}},
  \bibinfo{journal}{Phys. Rev. B} \textbf{\bibinfo{volume}{95}},
  \bibinfo{pages}{094511} (\bibinfo{year}{2017}),
  \urlprefix\url{https://link.aps.org/doi/10.1103/PhysRevB.95.094511}.

\bibitem[{\citenamefont{Wang et~al.}(2018)\citenamefont{Wang, Wang, and
  Sedrakyan}}]{Rui2018}
\bibinfo{author}{\bibfnamefont{R.}~\bibnamefont{Wang}},
  \bibinfo{author}{\bibfnamefont{B.}~\bibnamefont{Wang}}, \bibnamefont{and}
  \bibinfo{author}{\bibfnamefont{T.~A.} \bibnamefont{Sedrakyan}},
  \bibinfo{journal}{Phys. Rev. B} \textbf{\bibinfo{volume}{98}},
  \bibinfo{pages}{064402} (\bibinfo{year}{2018}),
  \urlprefix\url{https://link.aps.org/doi/10.1103/PhysRevB.98.064402}.

\bibitem[{\citenamefont{Wang et~al.}(2017)\citenamefont{Wang, Sedrakyan, Wang,
  and Xing}}]{Sedrakyan2017_12}
\bibinfo{author}{\bibfnamefont{R.}~\bibnamefont{Wang}},
  \bibinfo{author}{\bibfnamefont{T.~A.} \bibnamefont{Sedrakyan}},
  \bibinfo{author}{\bibfnamefont{B.}~\bibnamefont{Wang}}, \bibnamefont{and}
  \bibinfo{author}{\bibfnamefont{D.}~\bibnamefont{Xing}}
  (\bibinfo{year}{2017}), \urlprefix\url{https://arxiv.org/abs/1712.06762}.

\bibitem[{\citenamefont{Sedrakyan
  et~al.}(2015{\natexlab{b}})\citenamefont{Sedrakyan, Glazman, and
  Kamenev}}]{Sedrakyan2015}
\bibinfo{author}{\bibfnamefont{T.~A.} \bibnamefont{Sedrakyan}},
  \bibinfo{author}{\bibfnamefont{L.~I.} \bibnamefont{Glazman}},
  \bibnamefont{and} \bibinfo{author}{\bibfnamefont{A.}~\bibnamefont{Kamenev}},
  \bibinfo{journal}{Phys. Rev. Lett.} \textbf{\bibinfo{volume}{114}},
  \bibinfo{pages}{037203} (\bibinfo{year}{2015}{\natexlab{b}}),
  \urlprefix\url{https://link.aps.org/doi/10.1103/PhysRevLett.114.037203}.

\bibitem[{\citenamefont{Sohal et~al.}(2018)\citenamefont{Sohal, Santos, and
  Fradkin}}]{Sohal2018}
\bibinfo{author}{\bibfnamefont{R.}~\bibnamefont{Sohal}},
  \bibinfo{author}{\bibfnamefont{L.~H.} \bibnamefont{Santos}},
  \bibnamefont{and} \bibinfo{author}{\bibfnamefont{E.}~\bibnamefont{Fradkin}},
  \bibinfo{journal}{Phys. Rev. B} \textbf{\bibinfo{volume}{97}},
  \bibinfo{pages}{125131} (\bibinfo{year}{2018}),
  \urlprefix\url{https://link.aps.org/doi/10.1103/PhysRevB.97.125131}.

\bibitem[{\citenamefont{Bieri et~al.}(2015)\citenamefont{Bieri, Messio, Bernu,
  and Lhuillier}}]{Bieri2015}
\bibinfo{author}{\bibfnamefont{S.}~\bibnamefont{Bieri}},
  \bibinfo{author}{\bibfnamefont{L.}~\bibnamefont{Messio}},
  \bibinfo{author}{\bibfnamefont{B.}~\bibnamefont{Bernu}}, \bibnamefont{and}
  \bibinfo{author}{\bibfnamefont{C.}~\bibnamefont{Lhuillier}},
  \bibinfo{journal}{Phys. Rev. B} \textbf{\bibinfo{volume}{92}},
  \bibinfo{pages}{060407} (\bibinfo{year}{2015}),
  \urlprefix\url{https://link.aps.org/doi/10.1103/PhysRevB.92.060407}.

\bibitem[{\citenamefont{Bieri et~al.}(2016)\citenamefont{Bieri, Lhuillier, and
  Messio}}]{Bieri2016}
\bibinfo{author}{\bibfnamefont{S.}~\bibnamefont{Bieri}},
  \bibinfo{author}{\bibfnamefont{C.}~\bibnamefont{Lhuillier}},
  \bibnamefont{and} \bibinfo{author}{\bibfnamefont{L.}~\bibnamefont{Messio}},
  \bibinfo{journal}{Phys. Rev. B} \textbf{\bibinfo{volume}{93}},
  \bibinfo{pages}{094437} (\bibinfo{year}{2016}),
  \urlprefix\url{https://link.aps.org/doi/10.1103/PhysRevB.93.094437}.

\bibitem[{\citenamefont{Maiti and Sedrakyan}(2019)}]{Maiti2019}
\bibinfo{author}{\bibfnamefont{S.}~\bibnamefont{Maiti}} \bibnamefont{and}
  \bibinfo{author}{\bibfnamefont{T.~A.} \bibnamefont{Sedrakyan}},
  \bibinfo{journal}{arXiv} \textbf{\bibinfo{volume}{1812.10153}}
  (\bibinfo{year}{2019}), \urlprefix\url{https://arxiv.org/abs/1812.10153}.

\bibitem[{\citenamefont{Lu and Vishwanath}(2012)}]{Lu2012}
\bibinfo{author}{\bibfnamefont{Y.-M.} \bibnamefont{Lu}} \bibnamefont{and}
  \bibinfo{author}{\bibfnamefont{A.}~\bibnamefont{Vishwanath}},
  \bibinfo{journal}{Phys. Rev. B} \textbf{\bibinfo{volume}{86}},
  \bibinfo{pages}{125119} (\bibinfo{year}{2012}),
  \urlprefix\url{https://link.aps.org/doi/10.1103/PhysRevB.86.125119}.

\bibitem[{\citenamefont{Wen}(1995)}]{Wen1995}
\bibinfo{author}{\bibfnamefont{X.-G.} \bibnamefont{Wen}},
  \bibinfo{journal}{Advances in Physics} \textbf{\bibinfo{volume}{44}},
  \bibinfo{pages}{405} (\bibinfo{year}{1995}),
  \eprint{https://doi.org/10.1080/00018739500101566},
  \urlprefix\url{https://doi.org/10.1080/00018739500101566}.

\bibitem[{\citenamefont{Jo et~al.}(2012)\citenamefont{Jo, Guzman, Thomas,
  Hosur, Vishwanath, and Stamper-Kurn}}]{Jo2012}
\bibinfo{author}{\bibfnamefont{G.-B.} \bibnamefont{Jo}},
  \bibinfo{author}{\bibfnamefont{J.}~\bibnamefont{Guzman}},
  \bibinfo{author}{\bibfnamefont{C.~K.} \bibnamefont{Thomas}},
  \bibinfo{author}{\bibfnamefont{P.}~\bibnamefont{Hosur}},
  \bibinfo{author}{\bibfnamefont{A.}~\bibnamefont{Vishwanath}},
  \bibnamefont{and} \bibinfo{author}{\bibfnamefont{D.~M.}
  \bibnamefont{Stamper-Kurn}}, \bibinfo{journal}{Phys. Rev. Lett.}
  \textbf{\bibinfo{volume}{108}}, \bibinfo{pages}{045305}
  (\bibinfo{year}{2012}),
  \urlprefix\url{https://link.aps.org/doi/10.1103/PhysRevLett.108.045305}.

\bibitem[{\citenamefont{Atala et~al.}(2014)\citenamefont{Atala, Aidelsburger,
  Lohse, Barreiro, Paredes, and Bloch}}]{Atala2014}
\bibinfo{author}{\bibfnamefont{M.}~\bibnamefont{Atala}},
  \bibinfo{author}{\bibfnamefont{M.}~\bibnamefont{Aidelsburger}},
  \bibinfo{author}{\bibfnamefont{M.}~\bibnamefont{Lohse}},
  \bibinfo{author}{\bibfnamefont{J.~T.} \bibnamefont{Barreiro}},
  \bibinfo{author}{\bibfnamefont{B.}~\bibnamefont{Paredes}}, \bibnamefont{and}
  \bibinfo{author}{\bibfnamefont{I.}~\bibnamefont{Bloch}},
  \bibinfo{journal}{Nature Physics} \textbf{\bibinfo{volume}{10}},
  \bibinfo{pages}{588 EP } (\bibinfo{year}{2014}), \bibinfo{note}{article},
  \urlprefix\url{http://dx.doi.org/10.1038/nphys2998}.

\bibitem[{\citenamefont{Stenger et~al.}(1999)\citenamefont{Stenger, Inouye,
  Chikkatur, Stamper-Kurn, Pritchard, and Ketterle}}]{Stenger1999}
\bibinfo{author}{\bibfnamefont{J.}~\bibnamefont{Stenger}},
  \bibinfo{author}{\bibfnamefont{S.}~\bibnamefont{Inouye}},
  \bibinfo{author}{\bibfnamefont{A.~P.} \bibnamefont{Chikkatur}},
  \bibinfo{author}{\bibfnamefont{D.~M.} \bibnamefont{Stamper-Kurn}},
  \bibinfo{author}{\bibfnamefont{D.~E.} \bibnamefont{Pritchard}},
  \bibnamefont{and} \bibinfo{author}{\bibfnamefont{W.}~\bibnamefont{Ketterle}},
  \bibinfo{journal}{Phys. Rev. Lett.} \textbf{\bibinfo{volume}{82}},
  \bibinfo{pages}{4569} (\bibinfo{year}{1999}),
  \urlprefix\url{https://link.aps.org/doi/10.1103/PhysRevLett.82.4569}.

\bibitem[{\citenamefont{Rey et~al.}(2005)\citenamefont{Rey, Blakie, Pupillo,
  Williams, and Clark}}]{Rey2005}
\bibinfo{author}{\bibfnamefont{A.~M.} \bibnamefont{Rey}},
  \bibinfo{author}{\bibfnamefont{P.~B.} \bibnamefont{Blakie}},
  \bibinfo{author}{\bibfnamefont{G.}~\bibnamefont{Pupillo}},
  \bibinfo{author}{\bibfnamefont{C.~J.} \bibnamefont{Williams}},
  \bibnamefont{and} \bibinfo{author}{\bibfnamefont{C.~W.} \bibnamefont{Clark}},
  \bibinfo{journal}{Phys. Rev. A} \textbf{\bibinfo{volume}{72}},
  \bibinfo{pages}{023407} (\bibinfo{year}{2005}),
  \urlprefix\url{https://link.aps.org/doi/10.1103/PhysRevA.72.023407}.

\bibitem[{\citenamefont{Ernst et~al.}(2009)\citenamefont{Ernst, G{\"o}tze,
  Krauser, Pyka, L{\"u}hmann, Pfannkuche, and Sengstock}}]{Ernst2009}
\bibinfo{author}{\bibfnamefont{P.~T.} \bibnamefont{Ernst}},
  \bibinfo{author}{\bibfnamefont{S.}~\bibnamefont{G{\"o}tze}},
  \bibinfo{author}{\bibfnamefont{J.~S.} \bibnamefont{Krauser}},
  \bibinfo{author}{\bibfnamefont{K.}~\bibnamefont{Pyka}},
  \bibinfo{author}{\bibfnamefont{D.-S.} \bibnamefont{L{\"u}hmann}},
  \bibinfo{author}{\bibfnamefont{D.}~\bibnamefont{Pfannkuche}},
  \bibnamefont{and}
  \bibinfo{author}{\bibfnamefont{K.}~\bibnamefont{Sengstock}},
  \bibinfo{journal}{Nature Physics} \textbf{\bibinfo{volume}{6}},
  \bibinfo{pages}{56 EP } (\bibinfo{year}{2009}), \bibinfo{note}{article},
  \urlprefix\url{http://dx.doi.org/10.1038/nphys1476}.

\end{thebibliography}

\end{document}